\documentclass[3p,times,12pt]{elsarticle}

\journal{Nuclear Physics B}

\usepackage[colorlinks]{hyperref}
\usepackage{amsmath,amssymb}
\allowdisplaybreaks[2]

\usepackage{epstopdf}
\usepackage{amsmath}
\usepackage{amsfonts}
\usepackage{graphicx}
\usepackage{appendix}
\usepackage{dsfont}
\usepackage{amssymb}
\usepackage{bm}
\usepackage{color}
\usepackage{ulem}
\usepackage{soul}
\usepackage{natbib}
\usepackage{nccmath}
\usepackage{array}

\newcommand{\beq}{\begin{equation}}
\newcommand{\eneq}{\end{equation}}
\newcommand{\be}{\begin{equation}}
\newcommand{\ee}{\end{equation}}
\newcommand{\bea}{\begin{eqnarray}}
\newcommand{\eea}{\end{eqnarray}}

\usepackage{multirow}
\usepackage{wasysym}

\begin{document}

\begin{frontmatter}



\title{From four- to two-channel 
Kondo effect in junctions of XY spin chains }


\author[unical,infncos]{Domenico Giuliano\corref{cor1}}
\ead{domenico.giuliano@fis.unical.it}
\author[natal1,natal2]{Pasquale Sodano}
\ead{pasquale.sodano02@gmail.com}
\author[infncos,napoli,cnrspin]{Arturo Tagliacozzo }
\ead{arturo.tagliacozzo@na.infn.it}
\author[cnrdem,infntr]{Andrea Trombettoni}
\ead{andreatr@sissa.it}

\address[unical]{Dipartimento di Fisica, Universit\`a della Calabria,  
Arcavacata di Rende I-87036, Cosenza, Italy }
\address[infncos]{INFN, Gruppo collegato di Cosenza, 
Arcavacata di Rende I-87036, Cosenza, Italy }
\address[natal1]{International Institute of Physics, Universidade Federal 
do Rio Grande do Norte, 59078-400 Natal-RN, Brazil}
\address[natal2]{Departemento de F\'isica Teorica e Experimental,
Universidade Federal do Rio Grande do Norte, 59072-970 Natal-RN, Brazil}
\address[napoli]{Dipartimento di Fisica, Universit\`a di Napoli ``Federico II'', 
Monte S. Angelo-Via Cintia, I-80126 Napoli,Italy}
\address[cnrspin]{CNR-SPIN, Monte S. Angelo-Via Cintia, I-80126, Napoli, Italy}
\address[cnrdem]{CNR-IOM DEMOCRITOS Simulation Center, Via Bonomea 265, 
I-34136 Trieste, Italy}
\address[infntr]{SISSA and INFN, Sezione di Trieste, Via Bonomea 265, I-34136 
Trieste, Italy}
\cortext[cor1]{Corresponding author}

\begin{abstract}
We consider the Kondo effect 
in $Y$-junctions of anisotropic $XY$ models in an
applied magnetic field along the critical lines 
characterized by a gapless excitation spectrum. We find that, while the 
boundary interaction Hamiltonian describing the junction can be recasted 
in the form of a four-channel, spin-$1/2$ antiferromagnetic Kondo Hamiltonian, 
the number of channels effectively participating in the Kondo effect depends on
the chain parameters, as well as on the boundary couplings at the junction.
The system evolves from an effective four-channel topological Kondo 
effect for a junction of $XX$-chains with symmetric boundary couplings 
into a two-channel one at a junction of three quantum critical Ising chains. 
The effective number of Kondo channels depends on the properties of the 
boundary {\it and} of the bulk. The $XX$-line is a ''critical'' line, where a 
four-channel topological Kondo effect can 
be recovered by fine-tuning the boundary parameter, while along the line in 
parameter space connecting the 
$XX$-line and the critical Ising point the junction is effectively 
equivalent to a two-channel topological Kondo Hamiltonian. Using a 
renormalization group approach, we determine the flow of the 
boundary couplings, which allows us to define and estimate 
the critical couplings and Kondo temperatures  
of the different Kondo (pair) channels. Finally, we study the local 
transverse magnetization in the center of the $Y$-junction, 
eventually arguing 
that it provides an effective tool to monitor the onset of the two-channel 
Kondo effect.

\end{abstract}


\end{frontmatter}
 
\section{Introduction}
\label{intro}

The Kondo effect arises from the interaction
between magnetic impurities and itinerant electrons in a metal, resulting in a net low-temperature increase
of the resistance \cite{kondo64,hewson,kouwenhoven01}. Due to the large
amount of analytical and numerical tools developed to attack it and to
its relevance  for a variety of systems including heavy fermion materials \cite{hewson},
it became a paradigmatic example of a strongly interacting system and a
testing ground for new many-body techniques.

The Kondo effect has been initially studied for metals, like copper,
in which magnetic atoms, like cobalt, are added.
However, the interest in the Kondo physics
persisted also because it is possible to realize  it in a controlled way 
in devices such as, for instance, quantum dots
\cite{alivisatos96,kouwenhoven98}. Indeed,   when an odd number of   electrons is trapped
within the dot, it effectively behaves as a spin-1/2 localized magnetic
impurity: such a localized spin, placed
between two leads, mimics a magnetic impurity in a metal.

Another class of physical systems proposed to host the 
Kondo physics is provided by quantum spin chains \cite{eggert92,furusaki98,laflorencie08}:
the rationale is that the electron Kondo problem
may be described by a one-dimensional model since
only the $s$-wave part of the electronic wavefunction is affected by the
Kondo coupling. In \cite{laflorencie08} a magnetic impurity 
(i.e. an extra spin) at the end of a $J_1-J_2$ spin chain 
was studied and the comparison between the usual free electron Kondo model 
and this spin chain version was discussed: it was found that in general in the 
spin chain Kondo model  a  marginally irrelevant 
bulk interaction may be present, whose coupling 
in the $J_1-J_2$ model can be tuned to zero by choosing $J_2$ at the critical value of the 
gapless/dimerized phase transition. However, apart from the possible presence of these
marginal interactions modifying the details of the flow to the strong coupling Kondo fixed
point, the results indicate that the behavior is qualitatively the same as in 
the standard  Kondo effect  
\cite{laflorencie08,affleck09,bayat10,sodano10,deschner11,bayat12,tsve_2,bayat14,alkurtass15}. 
Moreover, it was also pointed out that 
a remarkable realization of the overscreened two-channel Kondo effect can be 
realized with two impurities in the $J_1-J_2$ chain  
\cite{bayat10,sodano10,bayat12,alkurtass15}. 

An important advantage of using spin chains to simulate 
magnetic impurities is that they provide in a natural 
way the possibility of nontrivially tuning the properties of the corresponding  low-energy effective Kondo Hamiltonian and to 
engineer  in a controllable way the impurity and its coupling 
to the bulk degrees of freedom. 
A recent paradigmatic example in this direction is provided by 
$Y$-junctions of suitably chosen spin chains \cite{crampettoni,tsve_1}. 
This kind of $Y$-junction 
is obtained when several (say, $M$) spin chains (the ''bulk'') 
are coupled to each other through a ''central region'' (the ''boundary''). 
For instance, one can 
couple the chains by connecting  the initial spins of each chain 
with each other 
with certain boundary couplings possibly different from the bulk ones.

Models of $Y$-junctions have been studied at the crossing, or the 
coupling, of three, or more, Luttinger liquids 
\cite{Kom,lal02,chamon03,giuliano08,agarwal14,mardanya15,giuliano15,berin1}, in 
Bose gases in star geometries \cite{Bur01,Bru04,Tok}, and in
superconducting Josephson junctions \cite{giuliano08,Cirillo}. 
Along this last
direction, it has also been proposed to simulate the two-channel Kondo effect at 
a $Y$-junction of quantum Ising chains \cite{tsve_1} 
or in a pertinently designed Josephson junction network \cite{giu_so_epl_2}.
A related system, a junction of $M$ classical two-dimensional 
Ising models, has been studied in \cite{fere}, 
where it was discussed the surface critical behavior, showing in particular 
that the $M \to 0$ limit 
corresponds to the semi-infinite Ising model in the presence of a 
random surface field.

In a $Y$-junction it is actually the coupling among bulk chains that 
determines the magnetic impurity. 
Formally, this arises from the extension of the 
standard Jordan-Wigner (JW) transformation
\cite{jordanwigner} to the non-ordered manifold provided by the 
junction of three (or more) chains \cite{crampettoni}. 
In order to preserve the correct 
(anti)commutation relations, this requires adding 
ancillary degrees of freedom associated 
to the central region, which is a triangle for $M=3$: 
after this JW transformation, that requires the introduction 
of appropriate Klein factors \cite{crampettoni},
the additional variables determine a spin variable magnetically 
coupled with the JW fermions from the chains which is topological, 
in view of the nonlocal 
character of the auxiliary fermionic variables \cite{BeriCooper2012}, 
realized as real-fermion Klein factors. 
On implementing this procedure 
in the case of a junction of three Ising chains in a transverse 
field tuned at their quantum critical point, one recovers 
a realization of the two-channel topological Kondo model \cite{tsve_1}. 
At variance, if one applies the procedure to a junction of three $XX$ chains, 
then one obtains  a realization of the four-channel topological Kondo model 
\cite{crampettoni}. In general, a reason of interest in studying such 
$Y$-junctions of spin chains is that it provides a remarkable physical 
realization of the topological Kondo effect 
\cite{BeriCooper2012,Altland2014multi,tsve_3,tsve_4,Buccheri2015,Buccheri15_bis,berin2}. 

In general, a {\it tunable} effective Kondo Hamiltonian can be realized 
at a $Y$-junction of quantum spin chains, provided the following requirements 
are met:

\begin{enumerate}

\item Of course, the spin chains should have tunable parameters; 
the solvability of the model on the chain is not strictly requested, 
but it helps in identifying  the correct mapping between the $Y$-junction 
of spin chains and the Kondo Hamiltonian;

\item The bulk Hamiltonian in the chains should be gapless 
(which is the case  of the critical Ising model in a transverse field \cite{tsve_1} 
and of the $XX$ model 
\cite{crampettoni}). Even though the study 
of boundary effects in $Y$-junctions of 
gapped chains may be interesting in its own, because of the competition  
between the scales given by the Kondo length and by  the 
correlation length associated to the bulk gap (similar to what 
happens for magnetic impurities in a superconductor \cite{AGD}, 
for the Josephson current 
in a junction containing resonant impurity levels \cite{glazman89} 
or for a quantum dot coupled to two superconductors 
\cite{campagnano,siano04,lim08}), yet, strictly speaking, in the case of a 
gapped bulk spectrum one does not recover a  Kondo fixed point.
\end{enumerate}

The $XY$ model in a transverse field which we consider 
in this paper meets both the above requirements in one shot, since:
\begin{itemize}
\item It is solvable for the uncoupled chains via JW transformations \cite{lieb61,korepin1997quantum,Takahashi}; 
\item It has two free parameters, the transverse field $H$ and the anisotropy $\gamma$ 
(in fact, the magnetic coupling $J$  can be just regarded as an over-all energy scale); 
\item Taken in pertinent limits, it reduces both to the $XX$ model 
($\gamma=1$, $H=0$) and to the critical Ising model 
($\gamma=0$, $H=2J$);
\item Last, but essential for our purposes, 
the parameters $H$ and $J$ can be chosen in a way that one can interpolate 
between the two limits, $XX$ and critical Ising, keeping a  
{\it gapless} bulk excitation spectrum.  
\end{itemize}  
 
After performing the JW transformation in the bulk and introducing the 
additional (''topological spin'') degrees of freedom at the junction, 
it has been established in \cite{crampettoni} that a 
junction of three quantum $XX$-chains hosts a spin-chain realization of 
the four-channel 
Kondo effect (4CK). 
At variance, in \cite{tsve_1} it has been shown that 
a junction of three critical quantum
Ising chains can be mapped onto a two-channel Kondo (2CK)-Hamiltonian. 
Mapping out the evolution of the system from the 4CK to the 2CK is 
the main goal of this paper: we eventually conclude that the 4CK effect of 
Ref.\cite{crampettoni} studied in the $XX$-point ($\gamma=1$, $H=0$) 
actually takes place along a ''critical'' line in 
parameter space, separating two 2CK systems, 
one of which is continuously connected to the 
2CK system corresponding to the Ising limit of \cite{tsve_1} 
by means of a continuous tuning of the 
bulk parameters of the junction. We also 
note that our reduction in the number of 
fermionic channels appears as the counterpart of the reduction in the 
''length'' of the  topological spin 
(from an $SO(M)$ to an $SO(M-2)$-vector operator)
by means of applied external fields, whose consequences are spelled out in 
detail in \cite{tsve_3}. In our case, the nature of the topological spin does 
not allow for defining local fields acting on it and, accordingly, we are
confined to an $SO(3)$ spin (which determines a non-Fermi liquid groundstate at
$T \to 0$ \cite{tsve_3}), with a number of effectively screening channels being 
either equal to $2$ or to $4$.
 
The plan of the paper is the following: 

\begin{itemize}
\item In section \ref{mapping_0} we write the model Hamiltonian 
for three anisotropic $XY$ chains in a transverse field 
on the $Y$-junction and 
we map it onto a spinless fermionic Hamiltonian by means of a pertinent 
JW-transformation;

\item We devote section \ref{cross.1} to study the transition 
from the four-channel Kondo to 
the two-channel Kondo regime, by continuously moving from the $XX$-line 
to the critical 
quantum Ising point, moving along  lines with a gapless bulk excitation spectrum;

\item In section \ref{kic} we study the behavior of the 
transverse magnetization at the junction as 
a function of the temperature and 
show how probing this can provide an effective mean
to monitor the onset of the Kondo regime;

\item Our conclusions and final comments are reported in section 
\ref{sec:concl}, while more technical material is presented in the appendices.

\end{itemize}

\section{The model and the mapping onto the Kondo Hamiltonian}
\label{mapping_0}

Our system consists of three 
$XY$-chains connected to each other 
via a 
boundary $XY$ interaction (the $Y$-junction),
involving the spins lying at one endpoint of 
each chain (we refer to these spins as the initial spins, as  
no periodic boundary conditions are assumed
in the chains). 
For the sake of simplicity, in the following we assume that the three 
chains are equal to each other, each one consisting of $\ell$ sites 
and with the 
same magnetic exchange interactions along the $x$- and the $y$-axis 
in spin space, respectively given by $-2 J$ and $- 2 \gamma J$ 
($J > 0 , 0 \leq \gamma \leq 1$), 
and the same applied magnetic field $H$ along the $z$-axis.
The Hamiltonian of the system $H_S$ is therefore given by 

 \beq
H_S = H_{XY} + H_\Delta\:\:\:\: . 
\label{xy.0}
\eneq
\noindent 
The ''bulk'' Hamiltonian $H_{XY}$ is given by the sum of three anisotropic $XY$ 
models in a magnetic field, that is, $H_{XY} = \sum_{ \lambda = 1}^3 H_{XY}^{(\lambda)}$,
with 

 \beq
 H_{XY}^{(\lambda)} = - 2 J \sum_{ j = 1}^{ \ell - 1 } 
\left(  S_{j , \lambda }^x S_{j+1 , \lambda }^x + \gamma S_{j , \lambda  }^y S_{j+1 , \lambda  }^y  \right)
  + H \sum_{ j = 1}^\ell S_{ j , \lambda }^z \:\:\:\: . 
\label{xy.1}
\eneq
\noindent
In Eq.(\ref{xy.1}), we denoted by ${\bf S}_{ j , \lambda } \equiv ( S^x_{ j , \lambda } , S^y_{ j , \lambda } , 
S_{j , \lambda }^z )$ the three components of a quantum spin-$1/2$ 
operator residing at site $j$ of the $\lambda$-th chain. 

At variance, the boundary Hamiltonian $H_\Delta$ in Eq.(\ref{xy.0}) is given by 

 \beq
H_\Delta = -   2 J_\Delta \: \sum_{ \lambda = 1}^3 \left( S_{1 , \lambda}^x S_{ 1 , \lambda + 1}^x + \gamma^{'} \: 
S_{1 , \lambda}^y S_{ 1 , \lambda + 1}^y \right)
\:\:\:\: , 
\label{xy.2}
\eneq
\noindent
with $\lambda + 3 \equiv \lambda$ (in other words, all the initial spins are 
coupled to each other by means of a magnetic exchange interaction $J_\Delta$ analogous to 
the bulk one, with anisotropy parameter equal to $\gamma'$). A possible 
additional contribution to $H_\Delta$ can be realized by means of 
a local magnetic field $H'$, as a term of the form 
$H' \sum_{ \lambda = 1}^3 S_{ 1 , \lambda }^z$.
Such a term may  be accounted for by locally modifying the applied magnetic 
field in the bulk Hamiltonian (\ref{xy.1}). For simplicity we will not consider it in the following since its inclusion does not qualitatively 
affect our conclusions.
 
By natural analogy with the assumption we make for the bulk parameters, 
in Eq.(\ref{xy.2}) we take $0 \leq \gamma' \leq 1$. Note that, in fact, this 
poses no particular limitations to the parameters' choice, since performing, 
for instance, on each spin a rotation by $\pi / 2$ along the $z$-axis in spin 
space allows for swapping the magnetic exchange 
interactions along the $x$- and the $y$-axis with each other. 
Also, in order for $H$ to be mapped on
a Kondo Hamiltonian at weak ''bare'' Kondo coupling, we assume 
a ferromagnetic boundary spin exchange amplitude  while, we pose no limitations 
on the sign of the bulk exchange amplitude, which is immaterial to 
our purpose, that is

 \beq
J_\Delta > 0; \:\:\:\: {\rm and} \:\:\:\: J_\Delta / \mid J \mid \leq 1 \:\:\:\: .
\label{xy.2bis}
\eneq
\noindent
From Eqs.(\ref{xy.1},\ref{xy.2}) one sees that for $\gamma = \gamma' = 1$ and 
$H=0$ the Hamiltonian (\ref{xy.0}) reduces back to the 
Hamiltonian for a star graph of three quantum $XX$-spin chains,
$H_{XX}$, studied in \cite{crampettoni}. At variance, for $\gamma = \gamma' = 0$,
it coincides with the Hamiltonian for the junction of three quantum Ising chains, 
introduced in \cite{tsve_1} and further discussed, together with  various generalizations, in 
\cite{tsve_2,tsve_3}. To map the $Y$-junction of $XY$-spin chains onto an effective 
Kondo Hamiltonian, we employ the standard JW-transformation 
\cite{jordanwigner,lieb61} generalized to 
a star junction of quantum spin chains \cite{crampettoni}. This eventually
allows us to resort 
to a pertinent description of the model in terms of spinless fermionic degrees of freedom. 

For a single, ''disconnected'' chain, the usual JW transformation \cite{jordanwigner,lieb61} consists on 
realizing the (bosonic) spin operators $S_j^z$ and 
$S^\pm_j$, where  $S^\pm_j=  S_j^x \pm i S^y_j   $, written 
in terms of spinless lattice operators $\{ c_ j, c_j^\dagger \}$ as 

 \begin{eqnarray}
 S_{ j  }^+ &=&  c_{ j  }^\dagger 
 e^{ i \pi \sum_{ r = 1}^{j-1} c_{ r  }^\dagger c_{ r  } }  \nonumber \\
 S_{ j  }^- &=&  c_{ j  } e^{ i \pi \sum_{ r = 1}^{j-1} c_{ r  }^\dagger 
 c_{ r  } }   \nonumber \\
 S_j^z &=& c_j^\dagger c_j - \frac{1}{2}
\;\;\;\; .
\label{xy.3}
\end{eqnarray}
\noindent
In \cite{crampettoni} it has been shown how, when 
generalizing Eqs.(\ref{xy.3}) to a junction of three quantum spin chains, besides 
adding to both spin and fermion operators the additional chain index $_\lambda$, in 
order to preserve the correct (anti)commutation relations, one has to 
add three additional Klein factors, that is, three real (Majorana) fermions
$\sigma^1 ,\sigma^2 , \sigma^3$, such that $\{ \sigma^\lambda ,\sigma^{ \lambda'} \} 
= 2 \delta_{ \lambda, \lambda'}$ (a  pictorial way of thinking about  
the $\sigma$-fermions is that they represent in the JW fermionic language 
the contribution of the junction degrees of freedom.) 
While a simple and effective way to consider $M>3$ 
has been presented in \cite{tsve_1}, in this paper we limit ourself 
to $M=3$, but we expect that 
the approach using the results of \cite{tsve_1} can be then 
generalized to $M>3$.

Once the additional Klein factors are introduced,
one generalizes Eqs.(\ref{xy.3}) to 

 \begin{eqnarray}
 S_{ j , \lambda}^+ &=& i c_{ j , \lambda}^\dagger e^{ i \pi \sum_{ r = 1}^{j-1} c_{ r , \lambda}^\dagger c_{ r , \lambda } } \:
\sigma^\lambda \nonumber \\
 S_{ j , \lambda}^- &=& i c_{ j , \lambda} e^{ i \pi \sum_{ r = 1}^{j-1} c_{ r , \lambda}^\dagger c_{ r , \lambda } } \:
\sigma^\lambda \nonumber \\
S_{j , \lambda}^z &=& c_{ j , \lambda}^\dagger c_{ j , \lambda} - \frac{1}{2}
\;\;\;\; . 
\label{xy.4}
\end{eqnarray}
\noindent
On implementing the generalized JW transformations in Eqs.(\ref{xy.4}), one
readily rewrites $H_{XY}$ as 
 
 \begin{eqnarray}
H_{XY} &=& - \frac{J ( 1 + \gamma ) }{2} \: \sum_{ \lambda = 1}^3 \;
\sum_{ j = 1}^{ \ell - 1} \left( c_{ j , \lambda}^\dagger c_{ j + 1 , \lambda } + 
c_{ j + 1 , \lambda}^\dagger c_{ j , \lambda } \right) \nonumber \\
&+&  \frac{J ( 1 -  \gamma ) }{2} \: \sum_{ \lambda = 1}^3 \;
\sum_{ j = 1}^{ \ell - 1} \left( c_{ j , \lambda} c_{ j + 1 , \lambda } + 
c_{ j + 1 , \lambda}^\dagger c_{ j , \lambda }^\dagger \right)
+ H \: \sum_{ \lambda = 1}^3 \;
\sum_{ j = 1}^{ \ell } c_{ j , \lambda}^\dagger c_{ j , \lambda} 
\:\:\:\: . 
\label{xy.5}
\end{eqnarray}
\noindent
From Eq.(\ref{xy.5}) one sees that, after the JW transformation, the Hamiltonian of 
each single chain  is mapped onto Kitaev Hamiltonian for a one-dimensional
$p$-wave superconductor \cite{kitaev}.

An important consistency check of the validity of  Eqs.(\ref{xy.4}) is that, as 
it must be, the Klein factors $\sigma^\lambda$ fully disappear from $H_{XY}$ in 
Eq.(\ref{xy.5}), which is the bulk Hamiltonian {\it without} 
the junction contribution. At variance,  the Klein factors play an important 
role in the boundary Hamiltonian $H_\Delta$ which, in fermionic coordinates, is 
given by 

 \beq
H_\Delta = 2 J_\Delta \, \left( \vec{\Sigma}_1 + \gamma' \vec{\Upsilon}_1 \right) 
\cdot \vec{\cal R}
\:\:\:\: .
\label{xy.6}
\eneq
\noindent
In analogy to \cite{crampettoni,tsve_1}, in Eq.(\ref{xy.6}) 
we have introduced the ''topological'' spin-$1/2$ operator 
$\vec{\cal R}$, whose components are bilinears of the Klein factors, 
defined as 

 \beq
\vec{\cal R} =  
- \frac{i}{2} \: \left( \begin{array}{c}
                       \sigma^2 \sigma^3 \\ 
\sigma^3 \sigma^1 \\ \sigma^1 \sigma^2                        
                      \end{array} \right)
\:\:\:\: . 
\label{xy.7}
\eneq
\noindent
The lattice operators  $\vec{\Sigma}_j , \vec{\Upsilon}_j$  are defined as 
 
 \beq
\vec{\Sigma}_j = 
- \frac{i}{2} \: \left( \begin{array}{c}
( c_{j,2} + c_{j,2}^\dagger)    ( c_{j,3} + c_{j,3}^\dagger) \\ 
( c_{j,3} + c_{j,3}^\dagger)    ( c_{j,1} + c_{j,1}^\dagger) \\ 
( c_{j,1} + c_{j,1}^\dagger)    ( c_{j,2} + c_{j,2}^\dagger)
       \end{array} \right) \;\;\; , \;\;  
\vec{\Upsilon}_j =  
 \frac{i}{2} \: \left( \begin{array}{c}
 ( c_{j,2} - c_{j,2}^\dagger)      ( c_{j,3} - c_{j,3}^\dagger)   \\ 
  ( c_{j,3} - c_{j,3}^\dagger)   ( c_{j,1} -  c_{j,1}^\dagger)   \\ 
  ( c_{j,1} -  c_{j,1}^\dagger)       ( c_{j,2} - c_{j,2}^\dagger)  
       \end{array} \right) 
\;\;\;\; . 
\label{xy.8}
\eneq
\noindent
As outlined in \cite{coleman} and reviewed in  \ref{spindensities}, 
both $\vec{\Sigma}_j $ and $\vec{\Upsilon}_j$ can be written as sum of 
two commuting lattice spin-$1/2$ operators. Therefore, $H_\Delta$ 
in Eq.(\ref{xy.6}) appears
to be associated to a four-channel, spin-$1/2$ Hamiltonian, 
with two pairs of channels coupled to 
$\vec{\cal R}$ with couplings respectively given by $J_1 = J_\Delta$ and $J_2 = \gamma' J_\Delta$: 
however, as it will be shown in the following, for values of $\gamma, J, H$ for which 
the bulk is gapless, the effective number of Kondo channels depends on the properties of the 
boundary {\it and} of the bulk. As $\gamma' = 1$, $H_\Delta$ becomes an isotropic 
four-channel Hamiltonian, consistently with the results of 
\cite{crampettoni} where the case $\gamma = \gamma' = 1$ and $H=0$ was considered. 

In the following we use Eqs.(\ref{xy.5},\ref{xy.6})
respectively for the bulk and the boundary Hamiltonian of the junction, to 
infer the boundary phase diagram of the junction, 
by particularly outlining the evolution of the system 
from the 2CK- to the 4CK-regime, and vice versa. Since, as   discussed in the Introduction, in 
order to recover an effective Kondo model 
one needs a gapless bulk excitation spectrum, 
we shall focus onto the critical, gapless lines of bulk Hamiltonian,  
along which we consider the crossover between a $Y$-junction of three $XX$-chains  
and the one of three  quantum Ising chains.

\section{The model along the gapless lines}
\label{cross.1}

The bulk spectrum of JW fermions at a $Y$-junction of 
quantum $XY$-chains in a transverse field is the same as for just a single chain,
which has been widely discussed in the literature  
\cite{korepin1997quantum,Takahashi,franchini}. The key result we need
for our purposes is the existence of  critical gapless lines in the space
of the bulk parameters $\gamma-H$. These are given by  

\begin{itemize}

\item the $XX$-line, corresponding to 
$\gamma = 1$, with $-2 J \leq H \leq 2 J$; 

\item the critical Kitaev line corresponding to $H =   J ( 1 + \gamma )$, 
 continuously connected to the endpoint $H = 2 J$ of the $XX$-line.
\end{itemize}

\begin{figure}
\includegraphics*[width=.6\linewidth]{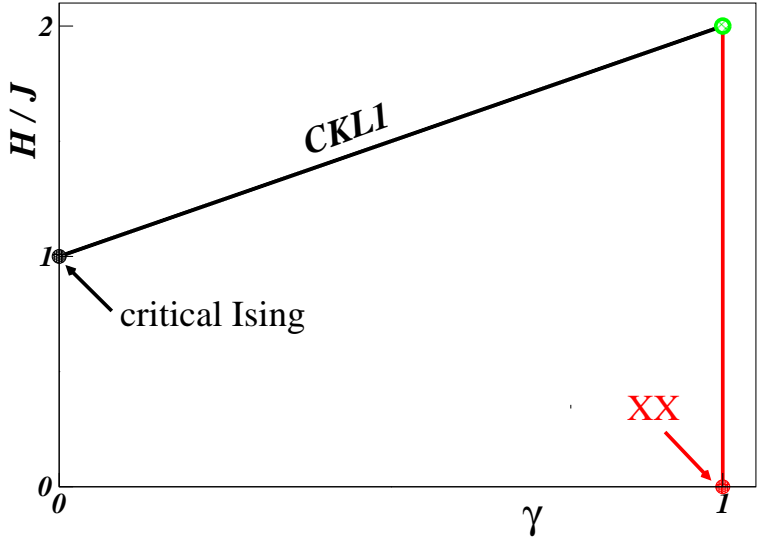}
\caption{Phase diagram of a single anisotropic 
$XY$-chain in a magnetic field in the $\gamma-H$ plane: the vertical red line at $\gamma=1$  
corresponds to the $XX$-line, while the black line corresponds to the 
CKL1  $H / J=1+\gamma$. The CKL1 intersects the $XX$-line at its endpoints, 
$(H , \gamma ) = ( 2 J , 1 )$, 
highlighted as a red filled black full dot. The opposite endpoint of CKL1, 
$(H , \gamma ) = ( J , 0 )$,
corresponds to a critical quantum Ising 
chain in a transverse magnetic field and is denoted by a black filled point. 
The merging of the two lines is also denoted by a green dot.} \label{critlines}
\end{figure}

In Fig.\ref{critlines}, we 
draw the critical lines in the $ \gamma-H$ plane for $H$ positive. 
The critical lines for $H<0$ are just symmetric to the ones plotted in 
Fig.\ref{critlines}. The limiting 
cases, evidenced in Fig.\ref{critlines}, are: {\it 1)} the critical Ising chain, where $H=J$ and $\gamma=0$ 
(where the gap is zero since the Ising chain is tuned at criticality);  
{\it 2)} the $XX$-chain, where $\gamma=1$ and $H=0$. 
The critical Kitaev line (CKL) in Fig.\ref{critlines} is denoted as CKL1: indeed there
is also a ''symmetric'' line, to which we refer to CKL2, corresponding to $ H = - J ( 1 + \gamma )$, 
continuously connected to the endpoint $H = - 2 J$ of the $XX$-line. Nevertheless, 
the behavior of the system along this latter line is the same as over 
the CKL. Therefore, 
we perform our analysis to the $XX$-line and mostly to CKL1, 
commenting as well on the CKL2. 

As we discuss in the following, the $XX$-line  is a ''critical'' line, where a 4CK can 
be recovered by fine-tuning the boundary parameter $\gamma' $ to 1. 
At its endpoint, such a line intersects the critical Kitaev line 
CKL1. As we are going to show next, along these lines the 
$Y$-junction can be effectively regarded as a 2CK-system. Therefore, 
CKL1 corresponds to an ''off-critical'' bulk displacement of the system, 
with a breaking of a 4CK Hamiltonian  down to a 2CK Hamiltonian, due to the 
breakdown of the {\it bulk} rotational symmetry in spin space. Other special points are 
the opposite endpoints of CKL1 and CKL2, corresponding to Tsvelik's $Y$-junction of
three critical quantum Ising chains \cite{tsve_1}.

In the following part of this Section, we implement a  
pertinently adapted version of Poor Man's 
renormalization group (RG) approach \cite{hewson}, 
 to infer under which conditions and in which form the Kondo 
effect is expected to arise at given values of the system's parameters. 
Specifically, we focus onto the main question of how the 
number of (JW) fermionic  channels screening the topological spin impurity
$\vec{\cal R}$ changes when moving along the phase diagram of Fig.\ref{critlines}.
Consistently   with the analyses of \cite{crampettoni,tsve_1}, 
one expects a change in the 
number of independent channels screening the topological spin impurity 
$\vec{\cal R}$. To explicitly spell this out, 
in the following  we also complement the perturbative results by 
performing a pertinent strong-coupling analysis of our system, by 
adapting to our specific case the regularization scheme of \cite{coleman}.

\noindent

\subsection{Poor Man's Renormalization Group}
\label{pmrg}

\begin{figure}
\includegraphics*[width=.6\linewidth]{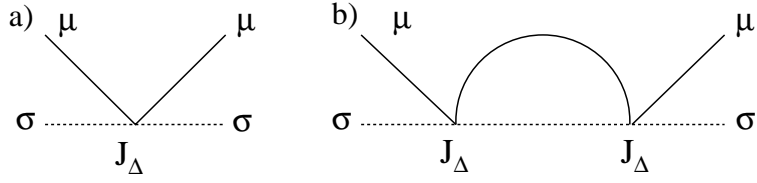}
\caption{Diagrammatic representation of the first- and second-order (in $J_\Delta$)
scattering processes arising from the perturbative expansion in a power
series of $S_\Delta$: {\it a)} First-order scattering vertex between $\mu$- and $\sigma$-fermions; 
{\it b)} Second-order loop process renormalizing the scattering amplitude to 
order $J_\Delta^2$. Notice that there is
only one loop diagram effectively renormalizing the coupling to second 
order in $J_\Delta$, at variance with what happens in the ''ordinary'' 
Kondo problem \cite{hewson}.} \label{diagram}
\end{figure}

To implement Poor Man's RG approach to the effective Kondo Hamiltonian 
at the junction, we begin by making a weak-interaction assumption for 
$H_\Delta$, which allows us to perturbatively account for the boundary 
interaction by referring to  the disconnected chain limit 
as a reference paradigm. To determine the energy eigenmodes and eigenfunctions in this limit, 
we consider the explicit diagonalization of a single anisotropic $XY$-chain with 
open boundary conditions (OBC), which we discuss in detail  in  \ref{solution_kitaev}.
The starting point is  the partition function, which we present  as 
$ {\cal Z } = {\cal Z}_0 \langle {\bf T}_\tau [ \exp{(- S_\Delta)} ] \rangle$, with 
${\bf T}_\tau$ being the imaginary time-ordered operator and the boundary 
action at imaginary times given by

 \beq
S_\Delta =  2 \sum_{ \lambda = 1}^3 \: \int_0^\beta \: d \tau \:
[ J_1 \Sigma_1^\lambda ( \tau )  + J_2  \Upsilon_1^\lambda ( \tau ) ] 
{\cal R}^\lambda ( \tau ) 
\;\;\;\; . 
\label{cro.1}
\eneq
\noindent
In Eq.(\ref{cro.1}) we have set $J_1 = J_\Delta , J_2 = \gamma' J_\Delta$. 
Moreover $\beta = ( k_B T)^{-1}$, $k_B$ being the Boltzmann constant (set to 
$1$ in the following for the sake of simplicity) and $T$ the 
temperature. In the standard RG approach to the Kondo problem, as the 
Kondo Hamiltonian represents a marginal boundary operator, typically one 
expands $\exp{(- S_\Delta)}$ up to second-order in $S_\Delta$ 
and then performs the appropriate two-fermion contractions, 
resulting in a cutoff-dependent 
correction to $J_1 , J_2$. This implies a 
logarithmic rise of the running couplings, due to the infrared divergences in 
the fermion propagators. To accomplish this point, it is useful to rewrite  
$S_\Delta$ in terms of the real $\mu$-fermions we introduce in 
 \ref{spindensities}, which are  related to the 
$c_{ j , \lambda}$ fermions by means of the relation
$c_{ j , \lambda} = \frac{1}{2} [ \mu_{ 2 j - 1 , \lambda} + i \mu_{2 j , \lambda} ]$.
The result is 

 \begin{eqnarray}
S_\Delta &=& - \sum_{ \lambda , \lambda' } \sum_{ j , j' = 1 , 2 } 
 \frac{ J_{ j , j'}}{2}  \: \int_0^\beta \: d \tau \: \sigma^\lambda ( \tau ) \sigma^{ \lambda' } 
( \tau ) \mu_{ j , \lambda } ( \tau ) \mu_{ j' , \lambda' } ( \tau ) \nonumber \\
 &=& \sum_{ \lambda } \sum_{ j , j' = 1 , 2 } 
2  J_{ j , j'} \frac{1}{\beta} \: \sum_\Omega {\cal R}^\lambda ( \Omega ) {\cal M}^\lambda_{ j , j' } 
(- \Omega ) 
\:\:\:\: . 
\label{cro.2}
\end{eqnarray}
\noindent
In Eq.(\ref{cro.2}) we have set $J_{1,1} = J_1$, $J_{2,2} = J_2$ and, in order to account for all 
the possible corrections arising from the second-order contractions, we introduced two 
additional effective couplings $J_{1,2},J_{2,1}$, which are set to $0$ in the bare
action. Moreover, on the second line we moved to the Matsubara-Fourier space 
and introduced the operator ${\cal M}_{ j , j'} ( \tau ) $ given by  

\beq
\vec{ \cal M}_{ j , j'} ( \tau ) = 
- \frac{i}{2} \left[ \begin{array}{c}
                      \mu_{ j , 2 } ( \tau ) \mu_{ j , 3 } ( \tau ) \\ 
                      \mu_{ j , 3 } ( \tau ) \mu_{ j , 1 } ( \tau ) \\ 
                      \mu_{ j , 1 } ( \tau ) \mu_{ j , 2 } ( \tau )  
                     \end{array} \right]
\:\:\:\: .
\label{ccro.4}
\eneq
\noindent
Incidentally, it is worth stressing here that the real fermions $\mu_{j , \lambda}$, with 
$j = 1 , 2$, physically reside at the same site (the first one) of the spin chains. 
Thus, the action $S_\Delta$ in Eq.(\ref{cro.2}) is a {\it pure boundary} one, 
as it only involves fermions at the first site of the chains. 
In order, now, to trade our perturbative derivation of the corrections to 
the coupling strengths for renormalization group equations in a ''canonical'' 
form we first of all assume that the temperature is low enough to trade 
the discrete sums over Matsubara frequencies for integrals. As per the 
standard derivation of Poor Man's scaling equations,
to regularize the integrals at high frequencies we introduce an 
ultraviolet (i.e., high-energy) cutoff $D$. This implies rewriting  Eq.(\ref{cro.2}) as  

\beq
S_\Delta  \approx \sum_{ \lambda } \sum_{ j , j' = 1 , 2 } 
2  J_{ j , j'} \int_{ - D}^D \: \frac{d \Omega }{2 \pi} \: 
{\cal R}^\lambda ( \Omega ) {\cal M}^\lambda_{ j , j' } 
(- \Omega ) 
\:\:\:\: . 
\label{ccro.b1}
\eneq
\noindent
In order to derive the RG equations for the running couplings, let us now
 rescale the cutoff from $D $ to $ D / \kappa$, with  
$0 < \kappa - 1 \ll 1$. In order to do this, we split the integral over $[ - D , D ]$ into an 
integral over $ [ - D / \kappa , D / \kappa ]$ plus integrals over  values 
of $\Omega$ within  $ [ D / \kappa , D ]$ and within $ [ - D , - D / \kappa ]$. In this latter integrals, 
we eventually integrate over the field operators. Therefore, from Eq.(\ref{ccro.b1}) we 
obtain, apart for a correction to the total free energy, which we do not consider here, as we
are mainly interested in the running coupling renormalization

\beq
S_\Delta \to  \int_{ - \frac{D}{\kappa}}^\frac{D}{\kappa}  \: \frac{d \Omega }{2 \pi} \: 
{\cal R}^\lambda ( \Omega ) {\cal M}^\lambda_{ j , j' } (- \Omega ) 
= \frac{1}{\kappa}  \int_{ - D }^D  \: \frac{d \Omega }{2 \pi} \: 
{\cal R}^\lambda \left( \frac{ \Omega }{ \kappa } \right) {\cal M}^\lambda_{ j , j' } 
\left(- \frac{ \Omega }{ \kappa } \right) 
\;\;\;\; , 
\label{ccc.1}
\eneq
\noindent
which, in view of the fact that $S_\Delta$ corresponds to a marginal boundary interaction
and, therefore, must be scale invariant, implies 

\beq
{\cal R}^\lambda \left( \frac{ \Omega }{ \kappa } \right) {\cal M}^\lambda_{ j , j' } \left(- \frac{ \Omega }{ \kappa } 
\right) 
= \kappa {\cal R}^\lambda ( \Omega ) {\cal M}^\lambda_{ j , j' } (- \Omega ) 
\:\:\:\: . 
\label{ccc.2}
\eneq
\noindent
Let us, now, look for second-order corrections to the running couplings. To this order, 
 we obtain a further correction $\delta S_\Delta^{(2)}$ to the first-order action by 
summing over intermediate states with energies  within  
$ [ D / \kappa , D ]$ and within $ [ - D , - D / \kappa ]$.
Due to the reality of the   $\mu_{j,\lambda}$'s, 
differently from what happens with the usual Kondo problem, to second order in $S_\Delta$,
there is only  one diagram effectively contributing to the corresponding renormalization of each
running coupling, which we draw in Fig.\ref{diagram}. Once the appropriate 
contractions have been done, the corresponding correction to $S_\Delta$ is given by

\noindent

\beq
\delta S_\Delta^{(2)}
 =   \sum_{ \lambda}   \sum_{ j , j' = 1 , 2 }  
2  J_{j,j} J_{j' , j '}
 \int_{ - D}^D \: \frac{d \Omega }{2 \pi} \: {\cal R}^\lambda ( \Omega ) {\cal M}^\lambda_{ j , j' } 
(- \Omega )  [ \Gamma^{ j , j' } ( D ; \eta ) +    \Gamma^{ j , j' } ( -  D ; \eta ) ] 
D ( 1 - \kappa^{-1} ) 
\:\:\:\: .
\label{ccro.3a}
\eneq
\noindent
with $\Gamma^{j , j' } ( \Omega  )$ being the Fourier-Matsubara transform of 
$\Gamma^{j , j' } ( \tau ) = G_{j , j } ( \tau ) g ( \tau )$,
the single-fermion Green's functions $G_{j , j } ( \tau )$ being listed in 
Eqs.(\ref{exp.18}), and 
$ g ( \tau ) = {\rm sgn} ( \tau )$ being the $\sigma$-fermion 
Green's function $g ( \tau ) = \langle {\bf T}_\tau [ \sigma ( \tau ) 
\sigma ( 0 ) ] \rangle$, that is

 \begin{eqnarray}
\Gamma^{ 1 , 1 } ( \Omega ) &=& \frac{1}{\beta} \: \sum_{\omega} 
 g ( \Omega  - \omega ) G^{ 1 ,1 } ( \omega ) \nonumber \\
  \Gamma^{ 2 , 2 } ( \Omega ) &=& \frac{1}{\beta} \: \sum_{\omega }  
 g ( \Omega - \omega ) G^{ 2 ,2 } ( \omega )  \nonumber \\
  \Gamma^{ 1 , 2 } ( \Omega ) &=& \frac{1}{\beta} \: \sum_{\omega }  
 g ( \Omega - \omega ) G^{ 1 ,2 } ( \omega_m )  
 \nonumber \\
   \Gamma^{ 2 , 1 } ( \Omega ) &=& \frac{1}{\beta} \: \sum_{ \omega }  
 g ( \Omega - \omega ) G^{ 2 ,1 } ( \omega )  
 \:\:\:\: . 
 \label{croma.4}
\end{eqnarray}
\noindent
When trading the sums over Matsubara frequencies for integrals, one has to introduce 
an infrared cutoff $\eta \sim  \pi T$ to 
cut the divergency  in $g  ( \Omega - \omega ) = 2 / ( i \Omega - i \omega )$
arising as $\omega \to \Omega$. Clearly, $\eta$ is related to the difference 
between fermionic and bosonic Matsubara frequencies, that is, 
$\eta \sim \pi T$. Accordingly, in Eq.(\ref{ccro.3a}) we evidenced the explicit dependence of 
$ \Gamma^{ j , j' } $ on both $\Omega$ and on the  cutoff $\eta$. In fact, rigorously performing the sum over the 
fermionic Matsubara frequencies already allows us to explicitly introduce the cutoff $\eta$,
at the price of retaining a parametric dependence of $\Gamma^{ j , j' }$ on $T$. 
We therefore obtain

 \begin{eqnarray}
\Gamma^{ 1 , 1 } ( \Omega ; \eta = \pi T ) &=&  - 2 
 \sum_{ \epsilon > 0 } \left[ \frac{ {\cal A}^2 ( \epsilon ) \epsilon }{\epsilon^2 + 
 \Omega ^2} \right] \tanh \left( \frac{\beta \epsilon}{2} \right) \nonumber \\
  \Gamma^{ 2 , 2 } ( \Omega ; \eta = \pi T )   &=&  = - 2
 \sum_{ \epsilon > 0 } \left[ \frac{ {\cal B}^2 ( \epsilon ) \epsilon }{\epsilon^2 + 
 \Omega^2} \right] \tanh \left( \frac{\beta \epsilon}{2} \right) \nonumber \\
  \Gamma^{ 1 , 2 } ( \Omega ; \eta = \pi T )    &=&     2 i 
 \sum_{ \epsilon > 0 } \left[ \frac{ {\cal A}  ( \epsilon ) {\cal B}  ( \epsilon ) 
 \Omega }{\epsilon^2 + 
 \Omega^2} \right] \tanh \left( \frac{\beta \epsilon}{2} \right) 
 \nonumber \\
   \Gamma^{ 2 , 1 } ( \Omega ; \eta = \pi T )   &=&   - 2 i 
 \sum_{ \epsilon > 0 } \left[ \frac{ {\cal A}  ( \epsilon ) {\cal B}  ( \epsilon ) 
 \Omega }{\epsilon^2 + 
 \Omega^2} \right] \tanh \left( \frac{\beta \epsilon}{2} \right) 
 \:\:\:\: . 
 \label{kkcro.4}
\end{eqnarray}
\noindent
 The explicit formula for the functions ${\cal A} ( \epsilon) , {\cal B} ( \epsilon )$ on 
the CKL1 is given in  Eqs.(\ref{exp.15}), while the one 
along the $XX$-line is given in Eq.(\ref{eee.1}). At the special points corresponding 
to a $Y$-junction of critical quantum Ising chains, they take the simplified 
expression in Eq.(\ref{exp.22}). Now, since    $\Gamma^{ 2 , 1 } ( \Omega ; \eta) $  
is an odd function of $\Omega$, we  get no renormalization for the 
$J_{1,2} , J_{2,1}$ that, therefore, get stuck at their bare values $J_{1,2} = J_{2,1} = 0$.
At variance, $J_{1,1} , J_{2,2}$ get renormalized to

 \begin{eqnarray}
 J_{1,1} &\to& J_\Delta - 2 J_\Delta^2 \Gamma^{1,1} ( D ; \eta  ) D ( 1 - \kappa^{-1} ) 
 = J_{1,1} - 2 J_{1,1}^2 \Gamma^{1,1} ( D ; \eta  )D ( 1 - \kappa^{-1} )  \nonumber \\
  J_{2,2} &\to& J_\Delta   \gamma'   - 2 J_\Delta^2 (\gamma' )^2 \Gamma^{2,2} ( D ; \eta  ) D ( 1 - \kappa^{-1} ) 
 = J_{2,2} -  2 J_{2,2}^2  \Gamma^{2,2} ( D ; \eta  ) D ( 1 - \kappa^{-1} ) 
\:\:\:\: . 
\label{cro.5}
\end{eqnarray}
\noindent
As a result, to leading order in 
$1 - \kappa^{-1} $,   we obtain that the variations of the 
running couplings as a function of $1 - \kappa^{-1}$ is given by 

\begin{eqnarray}
\delta J_{1,1} &=& -    J_{1,1}^2 \Gamma^{1,1} ( D ; \eta  )
D ( 1 - \kappa^{-1} ) \nonumber \\
\delta J_{2,2} &=& -    J_{2,2}^2 \Gamma^{2,2} ( D ;  \eta  )
D ( 1 - \kappa^{-1} ) 
\;\;\;\; .
\:\:\:\: . 
\label{kcro.6}
\end{eqnarray}
\noindent
From the perturbative result in Eqs.(\ref{kcro.6}) one works out the 
renormalization group equations for the running couplings $J_{ j ,j} ( D )$, 
with $D$ being the running scaling parameter. Specifically, one obtains 

\begin{eqnarray}
 \frac{d J_{1 , 1} ( D) }{d \ln \left( \frac{D_0}{  D } \right) }
&=& J_{1,1}^2 ( D ) \rho_{ 1 , 1 } ( D ;  \eta  ) \nonumber \\
 \frac{d J_{2 , 2} ( D ) }{d \ln \left( \frac{D_0}{ D } \right) }
&=&  J_{2,2}^2 ( T ) \rho_{ 2 , 2 } ( D ; \eta  )  
\:\:\:\: . 
\label{cro.6}
\end{eqnarray}
 \noindent
In Eqs.(\ref{cro.6}) we  have  denoted with  $D_0$  
a high-energy cutoff which we take equal to the upper-energy band 
edge, namely, $ D_0 =   2 J ( 1 + \gamma )  $, and have expressed the RG equations in terms of the functions
$\rho_{ j , j'} ( D ; \eta ) $ whose explicit form will be provided below in the various 
cases of interest. It is important to evidence, here, that, despite  their non-orthodox form, 
the RG equations in Eqs.(\ref{cro.6}) are expected to rigorously describe the scaling 
of the running couplings as a function of $D$. Indeed, we first of all notice the presence,
at the right-hand side of Eqs.(\ref{cro.6}), of the functions $\rho_{ j , j }$. This is typical
of Kondo  problems \cite{hewson} in the case in which one has an energy-dependent local density of 
states, (such as in the case of Kondo effect with superconducting leads \cite{campagnano,siano04,lim08}). Moreover,
we also note the  apparent additional dependence of   
the functions $\rho_{ j , j } ( D ; \eta )$ on $T$. In fact, this dependence enters just parametrically, 
through the  infrared cutoff $\eta$ and, at least at the initial cutoff rescaling step, is 
completely unrelated to the one on the scaling parameter $D$. At high enough temperature (that
is, when $T$ works as the ''natural'' infrared cutoff of the theory),  on rescaling the cutoff, the 
scaling trajectory is traversed from the initial scale $D_0$ and the initial 
coupling strengths $J_{j,j} ( D_0 )$, to an effective bandwidth $D_* \sim 2 \pi T$ and 
to effective running coupling strengths $J_{j,j} ( T )$, which become effectively 
temperature-dependent, whose dependence on $T$ can be deduced from integrating 
Eqs.(\ref{cro.6}) from $D=D_0$ to $D=2 \pi T$ \cite{hewson}. In fact, up to subleading contributions in $D^{-1}$, this is formally equivalent to solving 
the set of differential equations obtained by regarding the temperature $T$ as a running parameter,
which is given by 

\begin{eqnarray}
 \frac{d J_{1 , 1} ( T ) }{d \ln \left( \frac{D_0}{  T } \right) }
&=& J_{1,1}^2 ( T ) \rho_{ 1 , 1 } ( D = 2 \pi T ;  \eta = \pi T  ) \nonumber \\
 \frac{d J_{2 , 2} ( T ) }{d \ln \left( \frac{D_0}{ T } \right) }
&=&  J_{2,2}^2 ( T ) \rho_{ 2 , 2 } ( D = 2 \pi T ; \eta = \pi T )  
\:\:\:\: . 
\label{bello.1}
\end{eqnarray}
 \noindent
Eqs.(\ref{bello.1})  provide the reference differential system which we use in the following to
 determine the RG parameter flow in the various regimes of interest.
 Once the running couplings $J_{j,j} ( T )$ have been determined, we 
 resort to the effective ''renormalized'' boundary action by simply 
 substituting, in Eq.(\ref{cro.2}), $J_{j , j' }$ with $J_{ j , j'} ( T )$, that is 
 
 \beq
S_\Delta \to    \sum_{ \lambda } \sum_{ j , j' = 1 , 2 } 
2  J_{ j , j'} ( T )  \int_{ - D}^D \: \frac{d \Omega }{2 \pi} \: 
{\cal R}^\lambda ( \Omega ) {\cal M}^\lambda_{ j , j' } 
(- \Omega ) 
\;\;\;\; ,
\label{kcro.1bis}
\eneq
\noindent
with $J_{j,j} ( T )$ obtained by integrating Eqs.(\ref{bello.1}). 
In order to recast the formulas for  $J_{j,j} ( T )$ in a formula useful for our
further discussions, we follow Ref.\cite{lucada} in introducing  the 
''critical couplings'' $J_{j,j}^c$ defined as 

 \beq
J_{ j , j}^c \equiv \left\{ \int_0^{D_0} \: \rho_{ j , j } ( x ) \; \frac{d x }{x} \right\}^{-1}
\:\:\:\: , 
\label{cro.9}
\eneq
\noindent
in terms of which we obtain 

 \beq
J_{j , j } ( T ) = \frac{J_{j,j}^{(0)} J_{j,j}^c}{J_{j,j}^c - J_{j,j}^{(0)} + 
J_{j,j}^c J_{j,j}^{(0)} \int_0^T \: \rho_{ j , j } ( x ) \; \frac{d x }{x}}
\:\:\:\: , 
\label{cro.10}
\eneq
\noindent
with $J_{ j , j}^{(0)} = J_{j,j} ( D_0 ) $. Within Poor Man's RG approach, the onset
of the Kondo regime corresponds to the existence of a scale $T_K$ at which the 
denominator of Eqs.(\ref{cro.10}) is equal to $0$. Clearly, for channel-$j$ such a condition
can only met provided $J_{j,j}^{(0)} > J_{j,j}^c$. Notice that $j=1,2$, so, with 
"channel-$1$" we refer to the first pair of Konddo channels and with 
"channel-$2$" to the other pair.

By integrating Eqs.(\ref{bello.1}),  we now perform the perturbative RG analysis of the phase 
diagram of our $Y$-junction along the gapless lines in Fig.\ref{critlines}. 
In particular, the cornerstones of our discussion will be the endpoint of  
the CKL1 at $\gamma = 0$, where one recovers the junctions of three critical quantum 
Ising chains, discussed in \cite{tsve_1,tsve_2,tsve_3}, and the endpoint at 
$\gamma = 1$, shared with the $XX$-line. At the former point, we solve the 
RG equations and   provide the corresponding 
estimate for the Kondo temperature $T_K ( \gamma = 0 )$, which is consistent
with the formula presented in \cite{tsve_1}, once one   
resorts to consistent measure units for the temperature. At the latter point, 
we show that the 4CK Hamiltonian  at a junction of three $XX$-quantum spin chains which, 
when $- 2 J \leq H \leq 2J$, provides an extension of the specific system at $H=0$ 
discussed in \cite{crampettoni}, turns into an effective 
2CK Hamiltonian, due to a combined effect of the breaking of the rotational spin symmetry
in the bulk Hamiltonian $H_{XY}$, as well as in 
the boundary Kondo Hamiltonian $H_\Delta$. When moving along the CKL1, 
that is, when letting $\gamma$ continuously evolve from $\gamma = 1$ to $\gamma = 0$, 
we show that the system keeps hosting  two-channel Kondo effect. Therefore, 
from the qualitative point of view, any $\gamma < 1$ is equivalent to $\gamma = 0$, that is, 
at low temperatures/energies, the system flows towards the two-channel Kondo fixed point (2CKFP)
described in \cite{coleman}. In fact, while the explicit form of the 
leading boundary operator arising at the 2CKFP changes depending on whether $\gamma = 0$, 
or $0 < \gamma < 1$, there are no qualitative changes in the behavior of the system. 
Remarkably, this happens notwithstanding that, on increasing $\gamma$ towards $\gamma = 1$, 
at a critical value of  $\gamma$, $\gamma_c^{(2)}$, both couplings $J_1 ( T )  $ and $J_2 ( T )$ 
become relevant, as  $ J_1 ( T )$ crosses over towards the strong coupling regime before 
$J_2 ( T )$, due to the combined effect of having $\gamma < 1$ and  
$J_{1,1} ( D_0 ) > J_{2,2} ( D_0 )$.

\subsection{Onset of the Kondo regime along the critical Kitaev line}
\label{kkl}

We now discuss the perturbative RG approach to Kondo effect
along the CKL1. In fact, as it can be readily inferred from 
Eq.(\ref{xy.1}), reversing the sign of the bulk applied magnetic field $H$ is 
equivalent to performing the canonical uniform transformation

\beq
\left( \begin{array}{c}
        S_{j , \lambda}^x \\  S_{j , \lambda}^y \\  S_{j , \lambda}^z
       \end{array} \right) \longrightarrow \left( \begin{array}{c}
        S_{j , \lambda}^x \\ -  S_{j , \lambda}^y \\ - S_{j , \lambda}^z
       \end{array} \right) 
       \;\;\;\; , 
       \label{kkl.1}
\eneq
\noindent
that is, a uniform rotation by $\pi$ along the $x$-axis in spin space.
Therefore, our analysis can be readily extended to the symmetric CKL2 which 
we do not discuss in detail here. To recover the RG flow of the running couplings
$J_{1,1} ( T ) , J_{2,2} ( T )$, we need the explicit formulas for 
$\rho_{1,1} ( T ) = \rho_{1,1} ( D = 2 \pi T ; \eta = \pi T )$ and 
for  $\rho_{2,2} ( T ) = \rho_{2,2} ( D = 2 \pi T ; \eta = \pi T )$. In the large-$\ell$ limit,
these are given by

 \begin{eqnarray}
 \rho_{ 1,1} ( T ) &=& \frac{4}{ \pi T J ( 1 + \gamma) } 
 \int_\frac{1}{2 \pi}^\frac{J ( 1 + \gamma) }{\pi T} \: d u \: 
 \frac{u ( 1 - \delta )    \Sigma \left( \frac{2 \pi T u}{J ( 1 + \gamma ) }
 \right) }{   ( 1 + \delta ) ( 1 + u^2 )^2}
 \: e^{ \beta_q}  \tanh{( \pi u )}  \nonumber \\
  \rho_{ 2,2} ( T ) &=&   \frac{4}{ \pi T J ( 1 + \gamma) } 
 \int_\frac{1}{2 \pi}^\frac{J ( 1 + \gamma) }{\pi T} \: d u \: 
 \frac{u ( 1 +\delta )    \Sigma \left( \frac{2 \pi T u}{J ( 1 + \gamma ) }
 \right)  }{   ( 1 - \delta ) ( 1 + u^2 )^2}
 \: e^{-  \beta_q}  \tanh{(\pi u)}  
 \:\:\:\:, 
 \label{cro.7}
\end{eqnarray}
\noindent
with  $\delta = \frac{1 - \gamma}{1+\gamma}$, the integration variable $ u = \epsilon / ( 2 \pi T)$, 
the function $\Sigma ( x )$ given by

\beq
\Sigma ( x ) = \frac{\sqrt{2}}{1 - \delta^2}
 \Biggl\{ \sqrt{\delta^4 + ( 1 - \delta^2 )  x^2 }
 - \delta^2 - \frac{1 - \delta^2}{2}   x^2 \Biggr\}^\frac{1}{2}
\:\:\:\:,
\label{ccro.7}
\eneq
\noindent
and   $e^{ \pm \beta_q}  = \cosh ( \beta_q ) \pm \sinh ( \beta_q )$ 
are   defined from Eqs.(\ref{exp.11}) of  
\ref{solution_kitaev}. Along the CKL1 the critical couplings only 
depend on $\gamma$, that is, $J_{j,j}^c = J_{j,j}^c ( \gamma )$. 

\begin{figure}
\includegraphics*[width=0.7\linewidth]{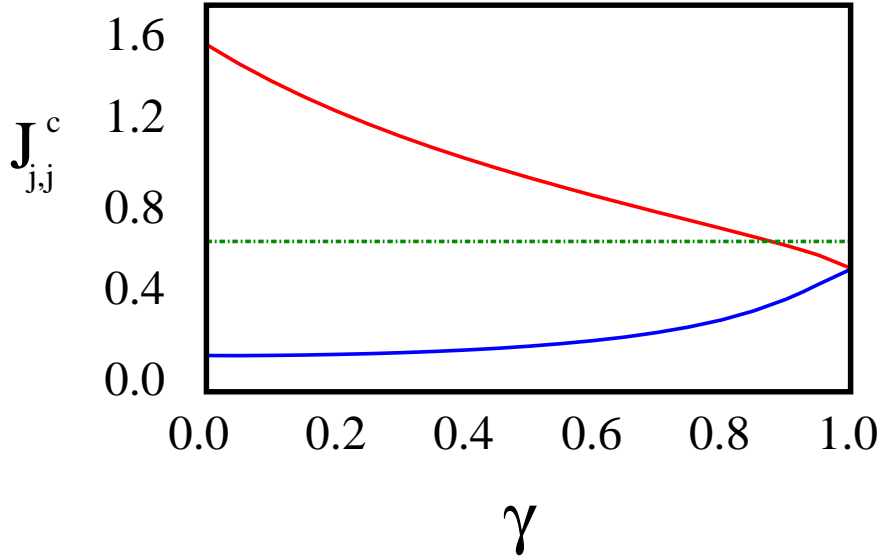}
\caption{Normalized critical couplings $J_{1,1}^c / [ J ( 1 + \gamma ) ] $ (blue curve) and  
$J_{2,2}^c  / [ J ( 1 + \gamma ) ]$
(red curve) as functions of $\gamma$. The dashed red line is a guide to the eye, corresponding 
to $J_{1,1}^{(0)}  / D_0=0.7$, which is the value we use to compute the Kondo temperature as a function 
of $\gamma$ for both channels.} 
\label{critcou}
\end{figure}

Using the explicit formulas for $\rho_{ j , j } ( T )$ in   Eqs.(\ref{cro.7}), 
from Eq.(\ref{cro.9})   we numerically evaluate 
$J_{j,j}^c ( \gamma ) / [ J ( 1 + \gamma ) ]$ for $0 \leq \gamma \leq 1$. 
The results are reported in Fig.\ref{critcou}. As it can be seen from the figure, 
$J_{1,1}^c ( \gamma ) / [ J ( 1 + \gamma ) ]$ keeps constantly different from zero and 
increases as $\gamma$ moves from $\gamma = 0$ to $\gamma = 1$, implying that, the closer
one gets to the isotropic $XX$-line, the harder is to develop Kondo effect in channel-$1$. On the 
other hand, $J_{2,2}^c ( \gamma ) / [ J ( 1 + \gamma ) ]$ increases as $\gamma$ increases 
towards 1,  till the curves for $J_{1,1}^c ( \gamma )  / [ J ( 1 + \gamma ) ]$
and for $J_{2,2} ( \gamma )^c  / [ J ( 1 + \gamma ) ]$ merge into each other, as $\gamma = 1$.
Thus, at variance with what happens in channel-$1$ (i.e., in the first 
pair of Kondo channels), the closer $\gamma$ 
is to 1, the easier for the system is to develop Kondo effect in 
channel-$2$ (i.e., in the other).  At the intersection between 
the CKL1 and the $XX$-line the two critical couplings are equal to each other and 4CK effect is 
recovered. To highlight what happens between the two endpoints of the CKL1, it is useful to 
think about it starting from the intersection point with the $XX$-line. This point belongs
to the critical line hosting an effective 4CK Hamiltonian. 4CK effect can be broken down to 2CK effect
by either breaking spin rotational invariance about the $z$-axis in the bulk, or in the boundary interaction Hamiltonian
(or in both of them). When moving to $\gamma < 1$ one is acting on the bulk spectrum: the breaking 
of the bulk spin rotational invariance about the $z$-axis suddenly implies a switch from 4CK- to 2CK-effect,
which persists and gets more and more robust, as long as one moves from $\gamma = 1$ towards the Ising 
limit $\gamma = 0$. The analogous breakdown of the spin rotational invariance about the $z$-axis  in 
$H_\Delta$ due to $\gamma' < 1$ just enforces the effect of the bulk spin anisotropy, but is not an 
essential ingredient, at least along the CKL1, where the system is effectively described in terms 
of an effective 2CK Hamiltonian, even at $\gamma' = 1$. 

\begin{figure}
\includegraphics*[width=0.6\linewidth]{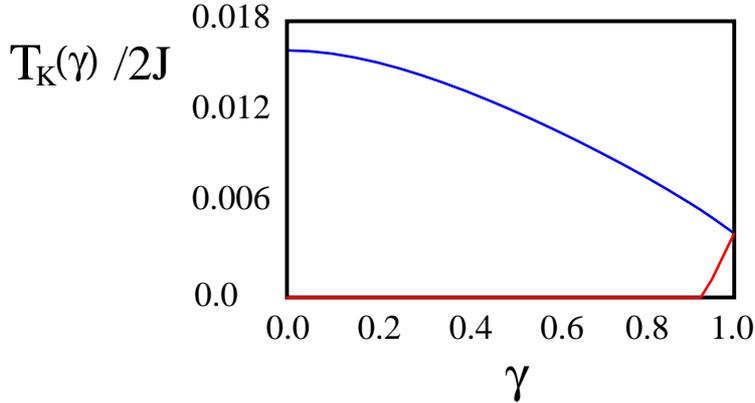}
\caption{Kondo temperature $T_{K  (1,1)} ( \gamma ) / J $ (blue curve) and 
$T_{K  (2,2)} ( \gamma ) /  J  $ (red curve), computed for 
$J_{1,1}^{(0)} / D_0  = 0.7$ and $J_{2,2}^{(0)}  = \gamma   J_{1,1}^{(0)} $ (see 
text). Consistently with the plot in 
Fig.\ref{critcou}, one sees that $T_{K   (2,2) } ( \gamma ) = 0$ for $0 \leq \gamma < 
\gamma_c$, with $\gamma_c \approx 0.92$. Also, one sees 
that $T_{K  (1,1)}  ( \gamma ) $ keeps constantly lower
 than $T_{K  (2,2)} ( \gamma ) $, implying a $T = 0$ fixed point 
 qualitatively equal to  the ''pure'' two-channel Kondo picture
 of the junction of three Ising chains \cite{tsve_1}. When $\gamma \to 1$ ($XX$-limit), 
 the  curves for $T_{K  (1,1)}  ( \gamma )$ and $T_{K  (2,2)}  ( \gamma )$ collapse 
 onto each other, consistently with the four-channel Kondo picture 
  of the junction of three $XX$-chains \cite{crampettoni}.} 
\label{kondotemp}
\end{figure}

To complement the result   emerging 
from the analysis of the critical couplings as functions of $\gamma$, 
we  numerically compute the Kondo temperatures $T_{K ; (1,1) } ( \gamma ),T_{K ; (2,2) } ( \gamma )$,
respectively associated to $J_{1,1} ( T ) , J_{2,2} ( T )$, as a function of $\gamma$ 
at given bare couplings $J_{1,1}^{(0)}  / D_0  = 0.7$ and 
$J_{2,2}^{(0)} / D_0 = \gamma  J_{1,1}^{(0)} / D_0 $ (which corresponds to 
 the ''natural choice'' $\gamma' = \gamma $). To do so, we start from  Eqs.(\ref{cro.10}) 
for $J_{1,1} ( T ) , J_{2,2} ( T )$ and 
identify $T_{K ( 1,1 )} ( \gamma ) $ [$ T_{K ( 2,2 )} ( \gamma ) $] as the 
value of $T$ at which the denominator of the formula for $J_{1,1} ( T )$ [$J_{2,2} ( T )$] becomes equal to 
$0$. The results are reported in Fig.\ref{kondotemp}.
First of all, consistently with the behavior of the
curves for $J_{1,1}^c ( \gamma )  / [ J ( 1 + \gamma ) ]$ and 
for $J_{2,2}^c ( \gamma )  / [ J ( 1 + \gamma ) ]$, we find that 
$T_{K; (1,1)} ( \gamma ) > T_{K; (2,2)} ( \gamma )  $ $\forall 
0 \leq \gamma < 1$. In particular, we find $T_{ K ( 1,1 )} ( \gamma = 0 ) \approx 0.016$, 
which is consistent with the result predicted using 
the formula for the Kondo temperature of a junction of three critical quantum
Ising chains provided in \cite{tsve_1} pertinently rewritten in 
our units, that is, $T_{K , {\rm Ising} } \sim \frac{J}{2 \pi} \exp \left(
- \frac{2 \pi J }{ 2 J_{1,1}^{(0)} } \right)$. 
In general, moving from $\gamma = 0$ to $\gamma = 1$, we find a remarkable
lowering of $T_{K (1,1)} ( \gamma )$. This can be roughly 
understood by recalling that the development of the Kondo effect
is mainly due to infrared divergences close to the Fermi level of 
itinerant fermions. We may, therefore, estimate $T_{ K (1,1) } ( \gamma )$ by 
pertinently approximating  $\rho_{1,1} ( T ) $ in Eqs.(\ref{cro.7}) (that is, by setting 
 $\tanh \left( \frac{z}{2 u} \right) \approx 1$ (which amounts to state that only 
the low-temperature regime matters in determining $T_K$), which leads
to the equation 

 \beq
\frac{ 1 - \delta }{ 1 + \delta } \:  \frac{2 J}{J_{1,1}^{(0)} } 
= \frac{ 16 \pi }{ ( 1 + \gamma ) }  \: \int_{ t_K}^\frac{1}{\pi} \: d u \: \int_0^2 \: d z \: 
\frac{u z \sin{k} }{( z^2 + 4 \pi^2 u^2 )^2} \: e^{ \beta_q} \: 
\:\:\:\: ,
\label{ckl.1}
\eneq
\noindent
with $t_K = T_{K   (1,1) }  / ( 2 J)$.  Eq.(\ref{ckl.1}) can be further simplified by noting that only 
the integration region around $z = 0$ is actually ''sensible'' to 
infrared divergences. Therefore, we approximate $\sin ( k )$ and 
$e^{ \beta_q }$ to leading order in $z$ and, at the same time, 
we introduce a cutoff $\lambda ( \gamma )$ to cut the integral 
over high values of $z$. As a result, we are able to further 
approximate Eq.(\ref{ckl.1}) as 

 \beq
\frac{ 1 - \delta }{ 1 + \delta }  \: \frac{2 J}{J_{1,1}^{(0)} } 
\approx \frac{2}{\pi ( 1 + \gamma ) } \: 
\frac{4 \delta^2}{1 - \delta^2} \int_0^{  \lambda  ( \gamma ) } 
\: d z \: \left( \frac{z}{z^2 + 4 \pi^2 t_K^2} - 
\frac{z}{z^2 + 4} \right)
\:\:\:\: . 
\label{ckl.2}
\eneq
\noindent
The cutoff in Eq.(\ref{ckl.2}) is chosen so that 
$ 2 \pi T_{K   (1,1) } ( \gamma ) / ( 2 J) \ll \lambda ( \gamma ) \ll 2$. On integrating 
Eq.(\ref{ckl.2}) one therefore obtains 

 \beq
T_{K   (1,1) }  ( \gamma ) \approx \frac{2 J ( 1 + \gamma )  \lambda^2 ( \gamma ) }{4 \pi} \: 
\exp \left[ - \frac{\pi 2 J  ( 1 + \delta )^2}{8 \delta J_{1,1}^{(0)} } \right] 
\:\:\:\: . 
\label{ckl.3}
\eneq
\noindent
As $\delta \to 1$ ($\gamma \to 0$), Eq.(\ref{ckl.3}) gives back the result for 
$T_{ K , {\rm Ising}}$, provided the cutoff is chosen so that 
$\lim_{ \gamma \to 0} \lambda ( \gamma ) = 1$. In general, comparing 
Eq.(\ref{ckl.3}) with the plot in Fig.\ref{kondotemp}, one may 
infer that $\lambda ( \gamma ) $ must be a smooth function of $\gamma$
around $\gamma = 0$ while, for $\gamma \to 1$, Eq.(\ref{ckl.3}) 
apparently fails to describe the numerically derived curve for $T_{ K (1,1)} ( \gamma )$. 
This is due to the fact that in this limit the density of states close to the Fermi level
is no more constant, but decreases as $| \epsilon |^\frac{1}{2}$, which 
makes less effective the effect of low-energy excitations close to the 
Fermi level and, therefore, less reliable  the approximation  
we performed. Finally, we note that, consistently with the result
drawn in Fig.\ref{critcou},  $T_{K (2,2)}  ( \gamma  ) = 0$
for $0 \leq \gamma \leq \gamma_c \approx 0.92$, which means that no 
Kondo effect develops in channel-$2$ for $ \gamma \leq \gamma_c$.  
  
\noindent

\noindent
We numerically checked that, changing the values of $J_{1,1}^{(0)} , J_{2,2}^{(0)}$ 
and alledging $\gamma ' \neq \gamma $, provides results only quantitatively different, 
but qualitatively similar, to what we draw in   Fig.\ref{kondotemp}. Therefore, we 
conclude that such a behavior of the Kondo temperatures associated to the two 
channels is quite a typical feature of the CKL1.

As a next step, we now infer what is  the $T = 0$ fixed point of the system along the  CKL1 by 
implementing the approach of \cite{coleman}, 
which we review and adapt to our specific case in  \ref{spindensities}.
To begin with, consistently with the 
perturbative RG analysis, we assume that, at a given $\gamma$, on lowering $T$ all the way down to 
$T_{K  (1,1)} ( \gamma )$, the system reaches a ''putative'' fixed point, where
the running coupling $J_{1,1} ( T )$ has crossed over towards the strongly-coupled
regime, while $J_{2,2} ( T )$ keeps finite (and perturbative). Consistently with the analysis of 
\cite{coleman}, this is a two-channel Kondo fixed point (2CKFP), with an
additional perturbation proportional to  the ''residual'' coupling $ J_{2,2} ( T_{K   (1,1) } )$. 
Therefore, we recover the two degenerate 
singlet ground states $ | \Sigma \rangle_1 , | \Sigma \rangle_2$ defined in 
Eqs.(\ref{sd.6}) and, as a leading perturbation at the 2CKFP, we 
obtain the operator $H_{\rm Sc}^{(2)}$ in Eq.(\ref{sd.11}) which, written in 
terms of the $c$-fermions, is given by 

 \begin{eqnarray}
H_{\rm Sc}^{(2)}  &=&   i \frac{J J_2}{2 J_1} {\bf V}^y \{  (1 + \gamma ) 
\prod_{ \lambda = 1, 2 , 3 } [ (-i ) ( c_{ 1 , \lambda } - 
c_{ 1 , \lambda}^\dagger ) ]  + i 
  \gamma [ ( c_{ 1 , 1 } - 
c_{ 1 , 1}^\dagger ) ( c_{ 2 , 2 } - 
c_{ 2 ,2}^\dagger )  ( c_{ 1 , 3 } - 
c_{ 1 , 3}^\dagger )  \nonumber \\
&+&  ( c_{ 1 , 1 } - 
c_{ 1 , 1}^\dagger ) ( c_{ 1 , 2 } - 
c_{ 1 ,2}^\dagger )  ( c_{ 2 , 3 } - 
c_{ 2 , 3}^\dagger )  + ( c_{ 2 , 1 } - 
c_{ 2 , 1}^\dagger ) ( c_{ 1 , 2 } - 
c_{ 1 ,2}^\dagger )  ( c_{ 1 , 3 } - 
c_{ 1 , 3}^\dagger ) ]   \} 
  \;\;\;\; , 
  \label{cro.a1}
  \end{eqnarray}
  \noindent
with the operator ${\bf V}^y$ directly connecting the two groundstates:
$ {\bf V}^y | \Sigma \rangle_{1 (2) } = - (+) i | \Sigma \rangle_{ 2 ( 1 ) }$
(see   \ref{spindensities} for details). $H_{\rm Sc}^{(2)} $ is 
a trilinear functional of fermionic operators. Therefore, it has 
scaling dimension $d_{\rm Sc}^{(2)} = \frac{3}{2} > 1$ and, accordingly, 
it is irrelevant at low energies/temperatures. A systematic Schrieffer-Wolff (SW) 
procedure, implemented by pertinently developing the formalism we 
use in   \ref{spindensities} to construct  $H_{\rm Sc}^{(2)} $, shows that
any other allowed boundary operator at 2CKFP-fixed point  is less relevant than
$H_{\rm Sc}^{(2)} $. As a result, we conclude that, since there are 
no relevant boundary operators arising at the  2CKFP, this is 
the actual  infrared stable fixed point of the system, along the 
whole CKL1.  Therefore, along this line the system is qualitatively
equivalent to the 2CK-system emerging at a junction of three quantum 
Ising chains \cite{tsve_1,tsve_2,tsve_3,tsve_4}, with the only difference that, in
this latter case,  
$H_{\rm Sc}^{(2)} $ is replaced with the third-order operator in 
Eq.(\ref{sd.13}). At variance, as we are going to discuss next, along 
the $XX$-line the $Y$-junction may either host 4CK-, or 2CK-physics, depending 
in this case only on the specific value of the boundary parameter $\gamma'$.

\subsection{Kondo effect along the $XX$-line}
\label{xxl}

The mapping of a  $Y$-junction of  three  XX quantum spin chains ($\gamma = 1$) 
at zero applied magnetic 
field ($H = 0$) onto an effective Kondo Hamiltonian has been discussed in 
\cite{crampettoni} by assuming symmetric couplings $J_{1,1}^{(0)} = J_{2,2}^{(0)} $
(that is, $\gamma' = 1$). As a  result, it has been found that such a system 
hosts a remarkable spin-chain realization of a 4CK-Hamiltonian. Here, we first  of all
show how the results of Ref.\cite{crampettoni} readily extend to the case $H \neq 0 , 
\gamma' = 1$, thus concluding that the whole $XX$-line, parametrized by $-2J \leq H \leq 2J$,
can be regarded as a ''critical'' 4CK-line, as long as $\gamma' = 1$. Therefore, we 
argue how a breakdown of the boundary coupling symmetry, that is, $\gamma' \neq 1$, 
corresponds to a relevant perturbation that lets the system flow towards a 2CKFP.
As a result, we conclude that the lack of rotational symmetry along the $z$-axis in 
spin space makes the system flow towards a 2CKFP, {\it even if} it is ''concentrated'' 
just at the boundary interaction Hamiltonian $H_\Delta$. 

\begin{figure}
\includegraphics*[width=0.7\linewidth]{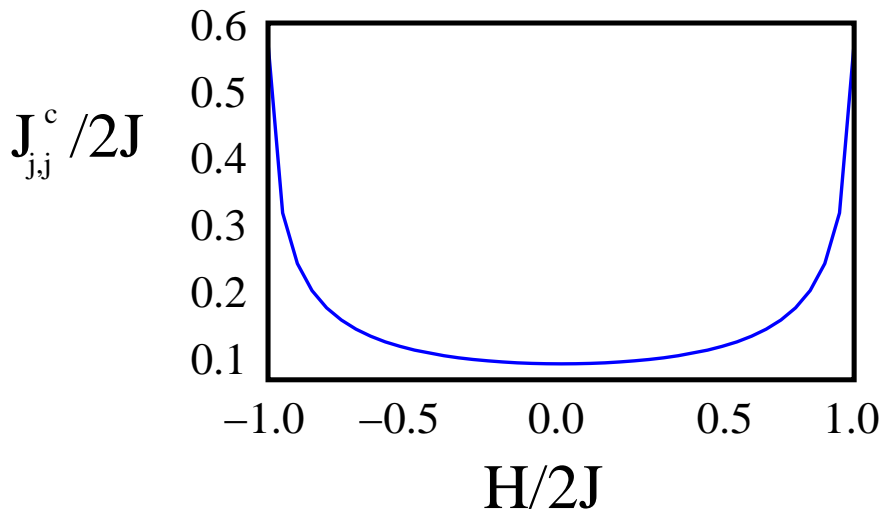}
\caption{$J_{j,j}^c ( h ) / ( 2 J ) $ {\it vs.} $h$ along the $XX$-line, 
parametrized by $- 1 \leq h \leq 1$. $J_{j,j} ( h ) $ is an even function of 
$h$, as it can be inferred from Eq.(\ref{xxo.2}). It is minimum at the 
point studied in Ref.\cite{crampettoni} ($h=0$) and, as $h \to \pm 1$, 
tends to the same value obtained by sending $\gamma$ to 1 along the CKL, that is, 
$J_{j,j}^c \approx 0.57  (2J)$.  } 
\label{j_xx}
\end{figure}

\begin{figure}
\includegraphics*[width=0.7\linewidth]{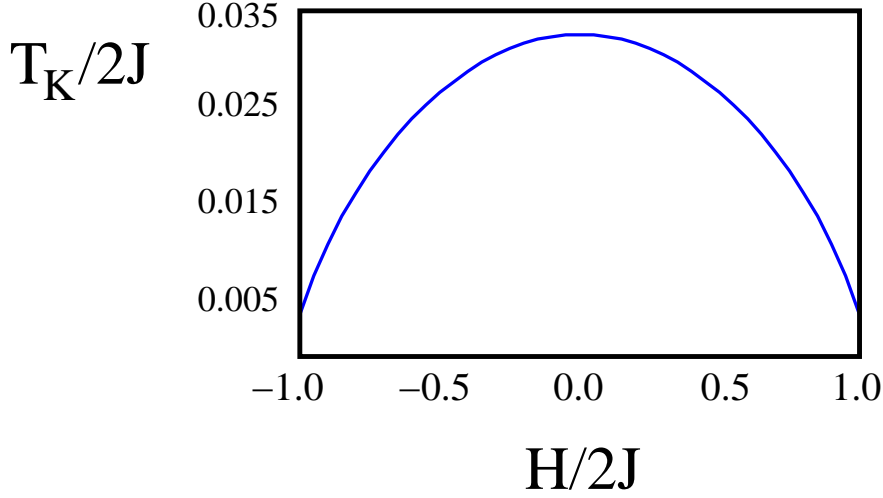}
\caption{Normalized Kondo temperature $T_K / (2 J )$ as a function of 
$h = H / ( 2 J)$ along the $XX$-line. The Kondo temperature is maximum 
at $h = 0$, as expected (see text), while, at the endpoints $h = \pm 1$, 
we obtain $T_K \approx 0.0041 (2J)$, which is the same value obtained 
by moving along the the CKL and towards the endpoint common to the $XX$-line
(that is, by sending $\gamma \to 1$ along the CLK). } 
\label{kondotemp_xx}
\end{figure}

To extend the discussion of the previous section to the $XX$-line, we 
note that, at a generic nonzero 
value of $H$, by direct calculation one obtains 

 \beq
\rho_{1,1} ( T )= \rho_{2,2 } ( T ) = 
\frac{16 \pi}{J } \: \int_{ - 2 J}^{ 2 J } \: d \epsilon \:
\: \frac{T^2 ( \epsilon - H ) \sqrt{1 - \left( \frac{\epsilon}{2 J} \right)^2} }{
 ( 4 \pi^2 T^2 + \epsilon^2 )^2 } \: \tanh \left( \frac{\epsilon - H}{2 T} 
 \right) 
 \:\:\:\:. 
 \label{xxo.1}
 \eneq
 \noindent
Using the definition in Eq.(\ref{cro.9}), we now evaluate the critical couplings
   $J_{1,1}^c ( h ) =  J_{2,2}^c ( h )$ as a function of  
$h = H / (2 J)$. As a result, we obtain 

 \beq
\frac{ 2 J}{ J_{1,1}^c } = \frac{ 2 J}{  J_{2,2}^c }  = 
16 \pi \: \int_0^\frac{1}{\pi} \: d x \: \int_{ -1}^1 \: d z \: 
\frac{ x ( z - h ) \sqrt{1 - u^2} }{ [ ( z - h )^2 + ( 2 \pi x )^2 ]^2 } \:
\tanh \left( \frac{z - h}{2 x} \right) 
\:\:\:\: , 
\label{xxo.2}
\eneq
\noindent
with $u = \epsilon / ( 2 J ), x = T / ( 2 J )$ (note that, due to the bulk 
spin rotational symmetry, that is, to the condition $\gamma = 1$, the 
condition $J_{1,1}^c ( h ) =  J_{2,2}^c ( h )$ is automatically recovered). 
In Fig.\ref{j_xx} we plot $J_{1,1}^c ( h ) / ( 2 J ) ( =  J_{2,2}^c ( h )  / ( 2 J ) $) 
as a function of 
$h$ for $- 1 \leq h \leq 1$. The plot is even for $h \to - h$, 
as it can be readily inferred from Eq.(\ref{xxo.2}). As $h \to \pm 1$, $J_{j,j}^c (h)$
tends to the same value obtained by sending $\gamma$ to 1 along the CKL1, that is, 
$J_{j,j}^c/(2J) \approx 0.57$. Apparently,  Fig.\ref{j_xx} shows that, the closer one
moves  towards the JW-fermion band edges ($h = \pm 1$), the harder is to 
recover Kondo effect. To double-check this conclusion, we also computed 
the Kondo temperature $T_K ( h )$ as a function of $h$ for $-1 \leq h \leq 1$,
by using the same technique we employed to compute $T_K ( \gamma )$ along 
the CKL1. We report the result of our numerical calculation for $T_K ( h )$ in 
Fig.\ref{kondotemp_xx}. As an over-all observation, we note that, as stated
above, the condition $\gamma = 1$ implies that $T_K ( h )$ is the same for both 
$J_{1,1} ( T )$ and for $J_{2,2} ( T )$ (that is the reason why we dropped the 
channel index from $T_K$). Moreover, just as $ J_{j,j}^c ( h )$  is an even function of $h$,
such is  $T_{K } ( h )$, as it appears from  Fig.\ref{kondotemp_xx}.

\noindent

\noindent
As for what concerns $T_K ( h )$,  we  see that, moving from either endpoint $h = \pm 1$ 
towards the center of the band ($h=0$), $T_K ( h )$ increases, till it reaches its maximum 
exactly at $h=0$. This is, in fact, consistent with the numerical estimate 
of  $J_{j,j}^c (h)$ reported in Fig.\ref{j_xx}, where one sees that 
 $J_{j,j}^c (h)$ is maximum at $h = \pm 1$ and decreases on moving from either endpoint 
towards the center of the band, till it reaches its minimum  exactly at $h=0$. 
Performing approximations analogous to the ones we did in section \ref{kkl}, 
the formula for $T_K$ along the $XX$-line can be presented as 

 \beq
 \frac{2 J}{J_{j,j}^{(0)} }   \approx  \frac{2}{\pi} \: \int_{-1 - h}^{1 - h} \: d z \: |z| \Biggl( 
\frac{1}{z^2 + 4 \pi^2 t_K^2} - \frac{1}{z^2 + 4  }  \Biggr) \sqrt{1 - (u + h)^2} \nonumber \\
\:\:\:\: ,
\label{xxlk.3}
\eneq
\noindent
with $t_K = T_K / ( 2 J)$. For $h=0$ the integral in 
Eq.(\ref{xxlk.3}) can be explicitly computed, yielding to the equation

 \beq
\frac{\pi J}{2 J_{j,j}^{(0)} } \approx - \ln ( \pi t_K ) - \sqrt{5} \:  {\rm atanh} \left( \frac{1}{\sqrt{5}} \right)
\:\:\:\: , 
\label{xxlk.4}
\eneq
\noindent
which, for $J_{j,j}^{(0)} / ( 2 J) = 0.7$, yields 

 \beq
T_K / (2J) \approx 0.035
\;\;\;\; , 
\label{xxlk.4_bis}
\eneq
\noindent
in good agreement with the numerical data.  For a generic value of $h$, 
assuming again that  the relevant contribution to the integral arises from the region
nearby $z = 0$, one may again introduce an ''ad hoc'' cutoff $\lambda ( h )$, so 
to approximate  Eq.(\ref{xxlk.3}) as 

 \beq 
\frac{2 J}{ J_{j,j}^{(0)}  }  \approx 
\frac{2}{\pi} \: \int_{- \lambda ( h ) }^{\lambda ( h ) } \: d u \: |u| \Biggl( 
\frac{1}{u^2 + 4 \pi^2 t_K^2} - \frac{1}{u^2 + 4  }  \Biggr) \sqrt{1 - h^2} 
\:\:\:\: , 
\label{xxlk.5}
\eneq
\noindent
with, again,  $ 2 \pi T_K ( h ) / ( 2 J )  \ll \lambda ( h ) \ll 2$.  
Eq.(\ref{xxlk.5}) is approximatively solved (for small $t_K$) by

 \beq
T_K ( h )  \approx  \frac{ \lambda^2 ( h )  J }{2 \pi} \: \exp \left[ - \frac{\pi 2 J }{ 4 J_{j,j}^{(0)}  \sqrt{1 - h^2} } 
\right]
\:\:\:\: , 
\label{xxlk.6}
\eneq
\noindent
with $\lambda ( h )$ being a smooth function of $h$ around $h=0$, such that 
$\lambda ( 0 ) = \exp \left[ -  \sqrt{5} \: {\rm atanh} \left( \frac{1}{\sqrt{5}} \right) \right] / \pi$.
Again, Eq.(\ref{xxlk.6}) fails to predict the finite value of $T_K ( h )$ as $h \to \pm 1$. 
The explanation is basically the same as for the limit $\gamma \to 1$, that is, 
in this limit the density of states close to the Fermi level
is no more constant, but decreases as $| \epsilon |^\frac{1}{2}$, which 
makes less effective the effect of low-energy excitations close to the 
Fermi level and, therefore, less reliable the approximation  
we performed. The important over-all conclusion is that, moving  from the case
$h = 0$ studied in \cite{crampettoni} does not qualitatively change the behavior of 
the system, but only quantitatively, as only $J_{j,j}^c $ and $T_k$ are affected.
Thus, $h (\in [ -1 , 1])$ appears to parametrize a line of points, all 
qualitatively equivalent to the one at $h=0$.

While the perturbative RG approach shows the (marginal) relevance of the
Kondo-like boundary interaction $H_\Delta$, it does not allow for an ultimate
characterization of the fixed point toward which the system flows as $T \to 0$.
To achieve this task, we have again to resort to a strong-coupling analysis
of the boundary interaction. For the sake of the discussion, it is more useful to 
assume    $J_{1,1}^{(0)} \neq J_{2,2}^{(0)}$. 
In this case, as it is readily inferred from the RG equations for 
$J_{1,1} ( T ) , J_{2,2} ( T )$ in Eqs.(\ref{cro.6}) and from Eq.(\ref{xxo.1})
for $\rho_{1,1} ( T ) , \rho_{2,2} ( T )$,  whichever coupling has the larger 
bare value (say $J_{1,1}$), reaches the strongly coupled regime first, at 
a Kondo temperature scale $T_{K , 1}$. At this scale, the two channels coupled
to $J_{1,1}$ develop 2CK effect with the topological spin $\vec{\cal R}$, while 
the two channels coupled to $J_{2,2}$ keep weakly
coupled to $\vec{\cal R}$, with effective coupling $J_{2,2} ( T_K )$.
Dubbing such a fixed point 2CKFP1, it is  possible to construct the leading 
boundary perturbation at 2CKFP1
along the procedure outlined in  \ref{spindensities}, exactly as we have done in section \ref{kkl}
for the CKL. The result is simply given by Eq.(\ref{cro.a1}) taken in the 
limit of $\gamma \to 1$, that is

 \begin{eqnarray}
H_{\rm Sc ; XX}^{(2)}  &=&   i \frac{J J_2  }{2 J_1} {\bf V}^y \{ 2
\prod_{ \lambda = 1, 2 , 3 } [ (-i ) ( c_{ 1 , \lambda } - 
c_{ 1 , \lambda}^\dagger ) ]  + i 
  [ ( c_{ 1 , 1 } - 
c_{ 1 , 1}^\dagger ) ( c_{ 2 , 2 } - 
c_{ 2 ,2}^\dagger )  ( c_{ 1 , 3 } - 
c_{ 1 , 3}^\dagger )  \nonumber \\
&+&  ( c_{ 1 , 1 } - 
c_{ 1 , 1}^\dagger ) ( c_{ 1 , 2 } - 
c_{ 1 ,2}^\dagger )  ( c_{ 2 , 3 } - 
c_{ 2 , 3}^\dagger )  + ( c_{ 2 , 1 } - 
c_{ 2 , 1}^\dagger ) ( c_{ 1 , 2 } - 
c_{ 1 ,2}^\dagger )  ( c_{ 1 , 3 } - 
c_{ 1 , 3}^\dagger ) ]   \} 
  \;\;\;\; . 
  \label{cro.a1xx}
  \end{eqnarray}
  \noindent
Again, the dimension counting yields for $H_{\rm Sc ; XX}^{(2)} $ a scaling dimension 
$d = \frac{3}{2}$, showing the irrelevance of this operator and, accordingly, the 
stability of the 2CKFP1. Despite the differences in the bulk spectrum, the 2CKFP1 
can be ''continuously deformed'' to the 2CKFP characterizing the junction along 
the CKL (at a given $h$ and $\gamma' = J_{2,2} ( D_0 ) / J_{1,1} ( D_0 )$, the
continuous deformation can be realized, for instance, by first moving to the 
intersection with the CKL1 (h=1) and, therefore, by further tuning $H$ and $\gamma$ 
keeping $H = - J ( 1 + \gamma )$, until  $\gamma = \gamma'$. 

At variance, along  the symmetric line, parametrized by $h$, corresponding to the symmetric initial condition 
$J_{1,1}^{(0)} = J_{2,2}^{(0)}$, the above argument does not apply. In this 
case, based on arguments similar to the ones used by 
Nozi\`eres and Blandin to infer the instability of Nozi\`eres 
Fermi liquid fixed point in the presence of overscreened Kondo effect
\cite{blandin}, one expects the system to flow, as $T \to 0$, to a novel fixed point,
different from the 2CKFP above. In fact,  this can be inferred by directly looking at the boundary Hamiltonian 
$H_\Delta$ in spin coordinates, in the limit $J_\Delta / J \to \infty$. To do so, let us 
generically write $H_\Delta$ as 

 \beq
H_\Delta = - J_1 \sum_{ \lambda = 1}^3 S_{1,\lambda}^x S_{1,\lambda + 1 }^x - 
J_2 \sum_{ \lambda = 1}^3 S_{1,\lambda}^y S_{1,\lambda + 1 }^y 
\equiv H_\Delta^{(1)} + H_\Delta^{(2)} 
\:\:\:\: . 
\label{bham.spin}
\eneq
\noindent
As discussed above, an asymmetric initial condition of the form, for instance, 
$J_1^{(0)} > J_{2}^{(0)}$ implies that the running coupling $J_1 ( T )$ reaches 
the strongly-coupled regime at the Kondo temperature $T_K$, at which 
$J_2 ( T_K )$ is still finite. At this point, one may attempt to recover 
the derivation of  \ref{spindensities} by using $H_\Delta^{(1)}$ as the unperturbed Hamiltonian 
and $H'$ given by $H' = H_\Delta^{(2)}  + H_T$, with 

\[
 H_T = - J \sum_{ \lambda = 1}^3 
\left( S_{1,\lambda}^x S_{2 , \lambda }^x + S_{1,\lambda}^y S_{2 , \lambda }^y \right)
\]
\noindent
as the perturbation. At $J_1$ large and positive, the groundstate manifold of 
$H_1$ is twofold degenerate, consisting of the degenerate, fully polarized states 
$ | \Leftarrow \rangle = | \leftarrow , \leftarrow , \leftarrow \rangle$ and
$ | \Rightarrow \rangle = | \rightarrow , \rightarrow , \rightarrow \rangle$, 
with $| \leftarrow \rangle $ and $ | \rightarrow \rangle$ being the two eigenstates of 
$S^x$. By direct calculation, one finds that all the matrix elements 
$ \langle X | H' | X' \rangle$, with $ | X \rangle , | X' \rangle = | \Leftarrow \rangle , 
| \Rightarrow \rangle$ are zero. As a result, the leading effective boundary 
Hamiltonian at the strongly-coupled fixed point is recovered to higher order in 
$H'$, by means of a SW procedure that realizes in spin coordinates the analog of what 
we have done in   \ref{spindensities} using JW fermionic coordinates. As 
we discuss in  \ref{spindensities} using JW fermions, this leads to an 
irrelevant operator, which shows the stability of the corresponding strongly 
coupled fixed point. At variance, in the symmetric case we construct a 
``putative'' fixed point by letting $J_1 / J  = J_2 / J \to \infty$. This 
leads to a twofold degenerate groundstate manifold, containing the 
states $ | \Uparrow \rangle , | \Downarrow \rangle$, given by

 \begin{eqnarray}
 | \Uparrow \rangle &=& \frac{1}{\sqrt{3}} \{ | \uparrow , \uparrow , \downarrow \rangle + 
 | \uparrow , \downarrow , \uparrow \rangle + | \downarrow , \uparrow , \uparrow \rangle \}
 \nonumber \\
 | \Downarrow \rangle &=& \frac{1}{\sqrt{3}} \{ | \downarrow , \downarrow , \uparrow \rangle + 
 | \downarrow , \uparrow , \downarrow \rangle + | \uparrow , \downarrow , \downarrow \rangle \}
 \:\:\:\: , 
 \label{disc.1}
\end{eqnarray}
\noindent
with $ | \uparrow \rangle , | \downarrow \rangle$ being the eigenstates of $S^z$. Clearly, in 
this case one obtains $H' = H_T$. By direct calculation one therefore checks that 
$ \langle \Uparrow | H' | \Uparrow \rangle = \langle \Downarrow | H' | \Downarrow \rangle =0$,
while one obtains 

 \begin{eqnarray}
 \langle \Downarrow | H' | \Uparrow \rangle &=& - \frac{2 J}{\sqrt{3}}
 \sum_{ \lambda = 1}^3 S_{ 2 , \lambda}^+ = - \frac{2 i J}{\sqrt{3}}
 \: \sum_{ \lambda = 1}^3 c_{ 2 , \lambda}^\dagger \sigma^\lambda 
 \nonumber \\
  \langle \Uparrow | H' | \Downarrow \rangle &=& - \frac{2 J}{\sqrt{3}}
 \sum_{ \lambda = 1}^3 S_{ 2 , \lambda}^- = - \frac{2 i J}{\sqrt{3}}
 \: \sum_{ \lambda = 1}^3 c_{ 2 , \lambda} \sigma^\lambda 
 \;\;\;\; , 
 \label{disc.2}
\end{eqnarray}
\noindent
with the right-hand side of Eqs.(\ref{disc.2}) containing both 
the spin and the JW-fermion representation of the leading 
boundary operators at the strongly-coupled fixed point. 
As a result, in this case, the leading boundary operator at the strongly 
coupled fixed point can be written as 

 \beq
H_{\rm Sc}^{(4)} =  - \frac{2 i J}{\sqrt{3}} \{ {\bf V}^- 
\sum_{ \lambda = 1}^3 c_{ 2 , \lambda}^\dagger \sigma^\lambda 
+ {\bf V}^+ 
\sum_{ \lambda = 1}^3 c_{ 2 , \lambda} \sigma^\lambda \}  
\:\:\:\: , 
\label{disc.3}
\eneq
\noindent
with the operators ${\bf V}^\pm$ defined, in analogy to ${\bf V}^y$ in 
Eq.(\ref{cro.a1}), so that ${\bf V}^+ | \Downarrow \rangle = 
| \Uparrow \rangle$,  ${\bf V}^- | \Uparrow \rangle = 
| \Downarrow \rangle$ and ${\bf V}^+ | \Uparrow \rangle = 
{\bf V}^- | \Downarrow \rangle = 0$. 
As Klein factors and ${\bf V}^\pm$ are non dynamical variables, 
both operators at the right-hand side of Eqs.(\ref{disc.3}) 
have scaling dimension $1/2$, which 
implies that $H_{\rm Sc}^{(4)}$ is a strongly relevant operator.
The putative fixed point $J_1 = J_2 = \infty$ is therefore not
a stable one. Based on arguments analogous to the ones
developed in \cite{crampettoni}, one expects that, in analogy with the
overscreened 2CK effect \cite{blandin}, at low temperatures/energies, 
the system flows towards an intermediate coupling fixed point, which 
  possibly corresponds to the  overscreened, 
spin-$1/2$ four-channel Kondo system \cite{aflud_1}.
Since any asymmetry between the bare couplings $J_{1,1} ( D_0 ) , J_{2,2} ( D_0 )$
makes the system flow towards a 2CKFP, we then conclude that the 
symmetric $XX$-line at $J_{1,1} ( D_0 ) =  J_{2,2} ( D_0 )$ is  a
critical 4CK-line marking, at fixed $h$, a quantum phase transition
between two 2CK-non-Fermi liquid phases. While it would be extremely 
interesting to discuss such a quantum phase transition in analogy to 
what is done, for instance, in \cite{pustilnik04}, where the overscreened 2CK-non-Fermi liquid
fixed point is regarded as a quantum phase transition between two perfectly
screened 1-channel Kondo Fermi-liquid phases, this  goes beyond the scope of this 
work and, accordingly, we plan to address this issue in a forthcoming publication. 
In the next section, instead, we discuss a possible tool to detect the onset 
of Kondo regime by means of an appropriate local magnetization 
measurement. 

\section{Kondo-induced crossover in the local transverse magnetization}
\label{kic}

The onset of Kondo effect is typically associated to a crossover 
in the effective dependence of running coupling strengths on a 
dimensionful scale, such as temperature, (inverse) length, etc., 
as one flows towards the Kondo fixed point: the dependence 
crosses over from the perturbative RG logarithmic rise, to 
a specific power-law scale, depending on the particular fixed point 
describing the system in the zero-temperature, large-scale limit. 
While in Kondo effect in metals and/or semiconducting devices
such a crossover is typically detected by looking at 
the current transport properties of the system as a function of the 
running scale, in our $Y$-junction of $XY$-chains
there are no free electric charges supporting electrical currents. 
Therefore, one has to resort to a different tool to monitor the 
onset of the Kondo regime. Specifically, we 
propose to look at the local transverse magnetization $m$ at the junction as a 
function of $T$, as $T$ is lowered all the way down towards $T_K$ and 
then to 0. We define $m$ as 
 \beq
m = \frac{1}{3} \: \sum_{ \lambda = 1}^3 \langle S_{1 , \lambda}^z \rangle 
\:\:\:\: .
\label{def_mag}
\eneq
\noindent
We refer to the quantity $m$ defined in Eq.(\ref{def_mag}) as the local 
transverse magnetization since the magnetic field $H$ in the Hamiltonian 
(\ref{xy.1}) is transverse with respect to the $x-y$ orientation of the spin 
(the order parameter of the phase transition for the Ising chain being 
then $\langle S^x \rangle $).

\begin{figure}
\includegraphics*[width=1.\linewidth]{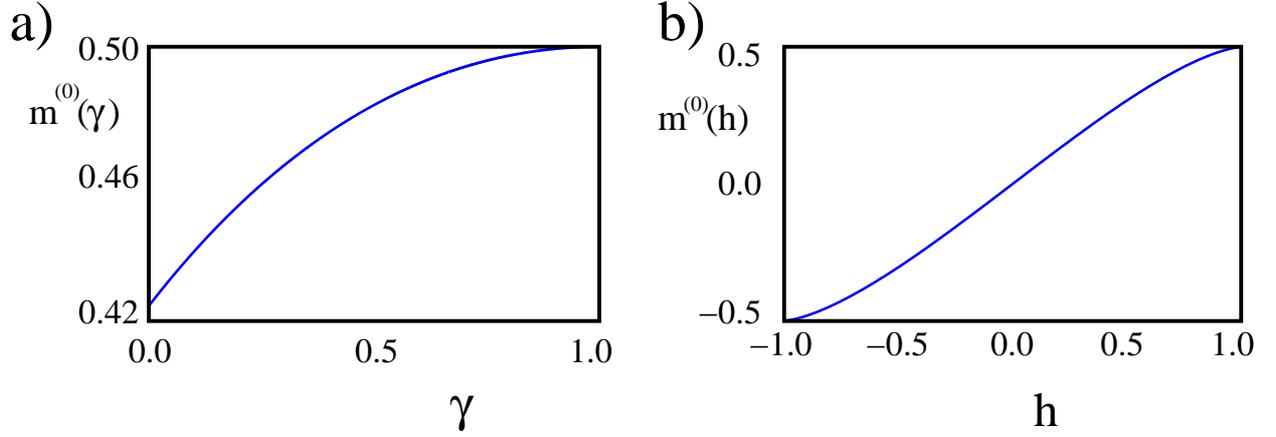}
\caption{{\bf a)}  Bare magnetization $m^{(0)} ( \gamma )$ along 
the CKL1 as a function of $\gamma$ for $0 \leq \gamma \leq 1$.
As $\gamma \to 0$, $m^{(0)} ( \gamma  \to 0)$ gives back the critical magnetization of 
the ferromagnetic quantum Ising chain, $m^{(0)} ( \gamma \to 0 ) \approx 0.424$, 
while $ m^{(0)} ( \gamma \to 1) \to 0.5$; 
{\bf b)} $m^{(0)} ( h )$ {\it vs.} $h$ along the $XX$-line ($-1 \leq h \leq 1$).
The line corresponds to the paramagnetic phase of the $XX$-model, in 
which $m^{(0)} ( h )$ is an odd function of $h$, and  $m^{(0)}  ( h \to \pm 1) \to 
\pm 0.5$, as it is appropriate at the paramagnetic-to-ferro(antiferro)magnetic 
phase transitions.} \label{mzero_bare}
\end{figure}

As a main motivation for our choice of $m$, we note that
it is difficult to implement a local magnetic
field acting on the topological spin, which led us to choose a 
different observable. 
In terms of the $\mu$-fermions, one can readily compute $m$ as 
 \beq
m = \frac{i}{6} \sum_{ \lambda = 1}^3 \langle \mu_{ 1 , \lambda } \mu_{ 2 , \lambda} 
\rangle = \frac{i}{6}  \lim_{ \eta \to 0^+ } 
\: \langle {\bf T}_\tau [  \mu_{ 1 , \lambda } ( \eta ) \mu_{ 2 , \lambda} ( 0 ) ]   
\rangle 
\:\:\:\: . 
\label{kic.3}
\eneq
\noindent
When the chains are disconnected from each other, there is no 
boundary effects and $m$  takes a finite, ''bare'' value  $m^{(0)}$, 
depending on  the bulk tuning parameters: $\gamma$ along the CKL1, and 
$h$ along the $XX$-line. Using  the Green's functions in Eqs.(\ref{exp.20}), 
  along the CKL1, one obtains 

\beq
m^{(0)}  ( \gamma )  =  \frac{1}{2 \pi J ( 1 + \gamma ) }
 \int_0^{ 2 J ( 1 + \gamma )  } \: d \epsilon  \: \Sigma \left( \frac{\epsilon}{J ( 1 + \gamma ) }
 \right)  
\:\:\:\: ,
\label{pertu.11}
\eneq
\noindent
with the function $\Sigma ( x )$ defined in Eq.(\ref{ccro.7}).
Along the $XX$-line, a similar calculation yields to an integral that 
can be explicitly calculated, giving the result

\beq
m^{(0)}  ( h )  = \frac{{\rm arcsin} ( h ) + h \sqrt{1 - h^2}}{ \pi}
\:\:\:\: . 
\label{pertu.14}
\eneq
\noindent
In Fig.\ref{mzero_bare}{\bf a)} we plot $m^{(0)}  ( \gamma ) $ {\it vs.} $\gamma$ 
computed along the CKL using Eq.(\ref{pertu.11}). The plot appears to be consistent 
with the known results for a single $XY$ model  in a 
transverse field \cite{lieb61,korepin1997quantum,Takahashi}, as highlighted in 
the figure caption. Similarly, in Fig.\ref{mzero_bare}{\bf b)} we plot $m^{(0)}  ( h ) $ 
{\it vs.} $h$ computed along the $XX$-line  using Eq.(\ref{pertu.14}). As $h \to 1$, 
we recover the value $0.5$ for  $m^{(0)}  ( h ) $, consistently with the numerical value 
obtained in the limit $\gamma \to 1$ for $m^{(0)}  ( \gamma ) $ along the CKL. Moreover, 
$m^{(0)}  ( h ) $ is an odd function of $h$, as expected from the properties of the 
Hamiltonian under a parity transformation in spin space. 

\begin{figure}
\includegraphics*[width=.7\linewidth]{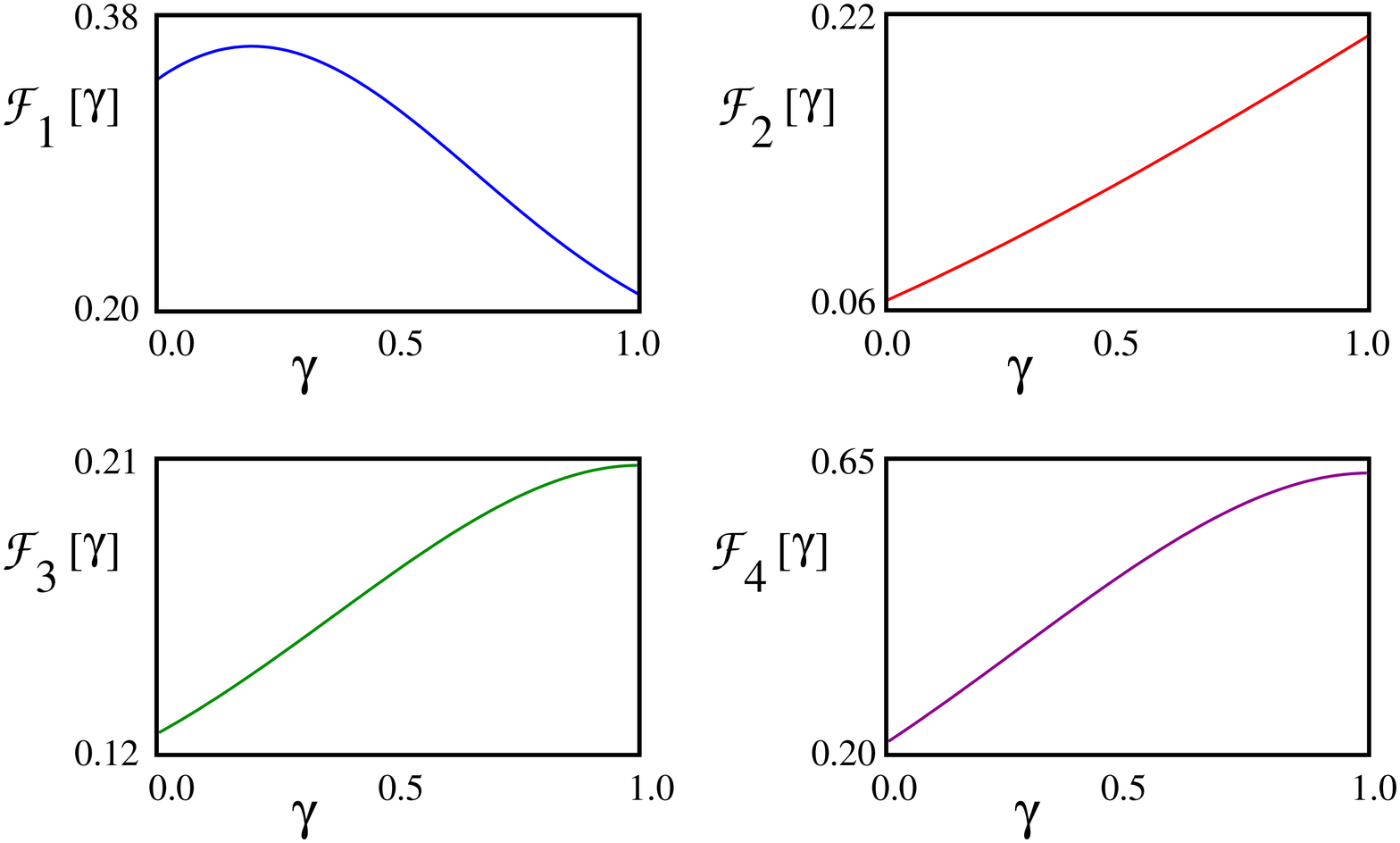}
\caption{Plot of the four functions ${\cal F}_j [ \gamma ]$ as a function of $\gamma$ for 
$0 \leq \gamma \leq 1$. In detail: {\it a) } ${\cal F}_1 [ \gamma ]$ for $0 \leq \gamma \leq 1$; 
{\it b) } ${\cal F}_2 [ \gamma ]$ for $0 \leq \gamma \leq 1$; 
{\it c) } ${\cal F}_3 [ \gamma ]$ for $0 \leq \gamma \leq 1$; 
{\it d) } ${\cal F}_4 [ \gamma ]$ for $0 \leq \gamma \leq 1$.
As it clearly appears from the plots, one obtains $\lim_{ \gamma \to 1} {\cal F}_1 [ \gamma ] = 
\lim_{ \gamma \to 1} {\cal F}_2 [ \gamma ] = \lim_{ \gamma \to 1} {\cal F}_3 [ \gamma ] \approx 0.21 \equiv 
{\cal F}_*$. 
At variance,  $\lim_{ \gamma \to 1} {\cal F}_4 [ \gamma ] \approx 0.63 = 3  {\cal F}_*$.
} \label{plot_delta}
\end{figure}

\begin{figure}
\includegraphics*[width=.7\linewidth]{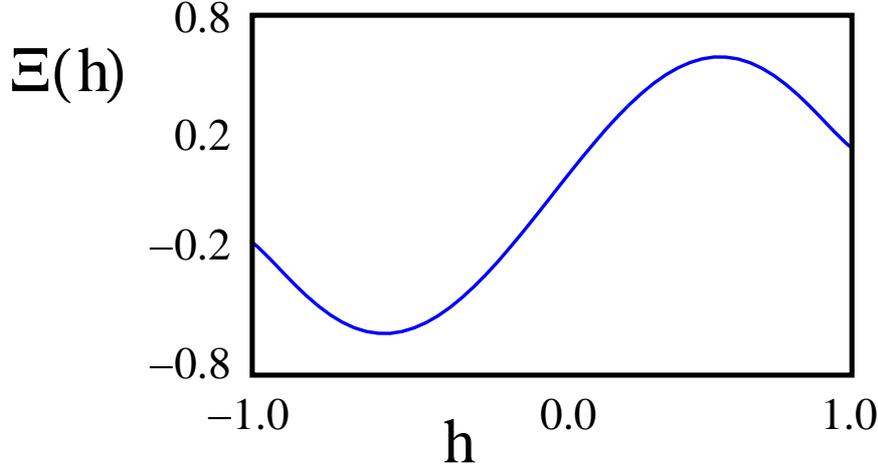}
\caption{Plot of $\Xi ( h )$ defined in Eq.(\ref{pertu.26}) as a function of $h$ for 
$-1 \leq h \leq 1$. As expected (see text), one obtains 
$\lim_{ h \to 1} \Xi ( h ) = {\cal F}_* \approx 0.21$.} 
\label{plot_h}
\end{figure}

When joining the chains to each other, the onset of the Kondo
regime induces an RG flow in the running couplings $J_{j,j} ( T )$.
Leaving, for the time being, aside the special symmetric case 
$J_{1,1}^{(0)} = J_{2,2}^{(0)}$ along the $XX$-line, in the previous section
we saw that, in all the other cases, the system flows towards a
strongly coupled fixed point in either coupling, say $J_{1,1}$, with the 
other coupling inducing a weak, irrelevant perturbation. Regarding the 
strongly coupled fixed point in spin coordinates, we see that, in order 
to minimize the energy, the system  lies within a ground state 
that is either one of the states $ | \Rightarrow \rangle , | \Leftarrow \rangle$
introduced above. Since    $ \langle \Leftarrow | \sum_{ \lambda = 1}^3 S_{1 , \lambda}^z  | 
\Leftarrow \rangle = \langle \Rightarrow | \sum_{ \lambda = 1}^3 S_{1 , \lambda}^z  | 
\Rightarrow \rangle = 0$, we find that, as a consequence of the onset of 
the Kondo regime,   $ m ( T )$ has to flow  to zero, 
as $T$ goes to zero. Moreover, knowing what are the leadingmost boundary operators
in the various regimes, allows us to infer the functional dependence of 
$m ( T ) $ on $T$ and, eventually, to interpolate the full crossover curve from 
$m^{(0)}$ all the way down to $ m ( T \to 0 )$. 
\noindent 
To investigate the effects of nonzero $J_{1  ,1 } $ and $J_{2 , 2 }$ on $m$ within perturbative 
RG approach, we first correct $m^{(0)}$ by means of a perturbative calculation in the boundary 
interaction strengths $J_1$ and $J_2$. Therefore, employing the  standard
approximations used in  the Kondo problem \cite{hewson}, after computing  
the correction $\delta m$ to the leading order in $J_{1 , 1}$ and $J_{2 , 2}$, we  
substitute the ''bare'' couplings, $J_{1 , 1}^{(0)} , J_{2 , 2}^{(0)}$, with  the running ones, 
$J_{1 , 1} ( T ) , J_{2 , 2} ( T )$, as derived in Eq.(\ref{cro.10}). 
To second-order in $J_{1 , 1} , J_{2 , 2}$ 
one obtains $m = m^{(0)} + \delta m$, with $\delta m$ given by

 \beq
\delta m = -  \lim_{ \eta \to 0^+} \frac{i}{8} \:
\sum_{ j  , j' = 1,2} J_{j , j} J_{j' , j'}  
\: \int_0^\beta \: d \tau_1 \: d \tau_2 \: 
 G_{ 1,j}( \eta - \tau_1 )  G_{2, j'} ( - \tau_2 ) 
G_{ j , j'} ( \tau_1 - \tau_2 )
\:\:\:\: . 
\label{pertu.21}
\eneq
\noindent
From  the expressions for the single-fermion Green's functions in 
Eqs.(\ref{exp.20}), in the $T \to 0$-limit, one obtains

 \begin{eqnarray}
 \delta m   &=& - 2  J_{1 ,1}^2  \:\sum_{ \epsilon_1 , \epsilon_2 , \epsilon_3 > 0 } {\cal A}^2 ( \epsilon_1 ) 
{\cal A} ( \epsilon_2 ) {\cal B} ( \epsilon_2 ) {\cal A}^2 ( \epsilon_3 ) 
\frac{  \epsilon_2}{  ( \epsilon_1 + \epsilon_2 ) ( \epsilon_1 + \epsilon_3 ) 
 ( \epsilon_2 + \epsilon_3 ) } \nonumber \\
 &-& 2  J_{2 , 2}^2  \:\sum_{ \epsilon_1 , \epsilon_2 , \epsilon_3 > 0 }
 {\cal A} ( \epsilon_1 ) {\cal B} ( \epsilon_1 ){\cal B}^2 ( \epsilon_2 )  
 {\cal B}^2 ( \epsilon_3 ) \frac{  \epsilon_1}{   ( \epsilon_1 + \epsilon_2 ) 
 ( \epsilon_1 + \epsilon_3 ) ( \epsilon_2 + \epsilon_3 ) } \nonumber \\
 &-& 2  J_1 J_2  \sum_{ \epsilon_1 , \epsilon_2 , \epsilon_3 > 0 } 
{\cal A}^2 ( \epsilon_1 )  {\cal B}^2 ( \epsilon_2 )  
{\cal A} ( \epsilon_3) {\cal B} ( \epsilon_3 ) 
\frac{  \epsilon_3}{ ( \epsilon_1 + \epsilon_2 ) ( \epsilon_1 + \epsilon_3 ) 
 ( \epsilon_2 + \epsilon_3 ) } \nonumber \\
&+&  2  J_{1 , 1} J_{2 , 2}  \sum_{ \epsilon_1 , \epsilon_2 , \epsilon_3 > 0 } 
{\cal A}  ( \epsilon_1 )  {\cal B}  ( \epsilon_1 )   {\cal A}  ( \epsilon_2 )  {\cal B}  ( \epsilon_2 ) 
{\cal A}  ( \epsilon_3 )  {\cal B}  ( \epsilon_3) 
\frac{ \epsilon_1 + \epsilon_2 +  \epsilon_3}{  ( \epsilon_1 + \epsilon_2 ) ( \epsilon_1 + \epsilon_3 ) 
 ( \epsilon_2 + \epsilon_3 ) } 
 \:\:\:\: ,
 \label{pertu.25}
\end{eqnarray}
\noindent
with ${\cal A} ( \epsilon ) ,  {\cal B} ( \epsilon )$ defined in Eqs.(\ref{exp.15}).
Regarding, as stated above, $J_{1,1} , J_{2,2}$ as running couplings along 
the CKL1, we obtain  

 \beq
\delta m ( T ; \gamma )  =  -   2 J_{1 ,1}^2 ( T )  \: {\cal F}_1 [ \gamma ] 
-  2 J_{2 , 2}^2 ( T )   \: {\cal F}_2 [ \gamma ] - 
 2 J_{1 , 1}  ( T ) J_{2 , 2}   ( T )  \: \{  {\cal F}_3 [ \gamma ] 
- {\cal F}_4 [ \gamma ] \}
\:\:\:\: , 
\label{pertu.29}
\eneq
\noindent
with 

 \begin{eqnarray}
&& {\cal F}_1 [ \gamma  ] = 
 \frac{1}{2} \: \gamma^2 
 \: \left( \frac{2}{\pi} \right)^3 \prod_{ j = 1}^3 \left( 
 \int_0^{ 2 J ( 1 + \gamma )  }  \frac{d \epsilon_j  }{  J ( 1 + \gamma )  } 
 \: \Sigma \left( \frac{\epsilon_j}{
 J ( 1 + \gamma) } \right)\right) 
 \left[ 
 \frac{  \epsilon_2  
 e^{ \beta_{q,1} } e^{  \beta_{q , 3} }}{ ( \epsilon_1 + \epsilon_2 ) ( \epsilon_1 + \epsilon_3 ) 
 ( \epsilon_2 + \epsilon_3 ) } \right]  
 \nonumber \\
 &&  {\cal F}_2 [ \gamma  ] =  
 \frac{1}{2} \: \gamma^{-2} 
 \: \left( \frac{2}{\pi} \right)^3 \prod_{ j = 1}^3 \left( 
 \int_0^{ 2 J ( 1 + \gamma )  }  \frac{d \epsilon_j  }{   J ( 1 + \gamma )  }
 \: \Sigma \left( \frac{\epsilon_j}{
 J ( 1 + \gamma) } \right) \right) \left[ 
 \frac{  \epsilon_1  
 e^{ - \beta_{q,2} } e^{  -\beta_{q , 3} }}{ ( \epsilon_1 + \epsilon_2 ) ( \epsilon_1 + \epsilon_3 ) 
 ( \epsilon_2 + \epsilon_3 ) } \right]   ]
 \nonumber \\ 
  &&  {\cal F}_3 [ \gamma ] =  
 \frac{1}{2}
 \: \left( \frac{2}{\pi} \right)^3 \prod_{ j = 1}^3 \left( 
 \int_0^{ 2 J ( 1 + \gamma )  }  \frac{d \epsilon_j  }{   J ( 1 + \gamma )  } 
 \: \Sigma \left( \frac{\epsilon_j}{
 J ( 1 + \gamma) } \right) \right) 
 \left[ \frac{  \epsilon_3 
 e^{  \beta_{q,1} } e^{  -\beta_{q , 2} } }{ ( \epsilon_1 + \epsilon_2 ) ( \epsilon_1 + \epsilon_3 ) 
 ( \epsilon_2 + \epsilon_3 ) }  \right]   
 \nonumber \\
&&{\cal F}_4 [ \gamma ] =   
 \frac{1}{2} \: 
 \: \left( \frac{2}{\pi} \right)^3 \prod_{ j = 1}^3 \left( 
 \int_0^{ 2 J ( 1 + \gamma )  }  \frac{d \epsilon_j  }{   J ( 1 + \gamma )  }
 \: \Sigma  \left( \frac{\epsilon_j}{
 J ( 1 + \gamma) } \right) \right) 
 \left[ \frac{ ( \epsilon_1 + \epsilon_2 +  \epsilon_3 ) 
 }{ ( \epsilon_1 + \epsilon_2 ) ( \epsilon_1 + \epsilon_3 ) 
 ( \epsilon_2 + \epsilon_3 ) }  \right]  
 \:\:\:\: . 
 \label{pertu.28}
\end{eqnarray}
\noindent
At variance, along the $XX$-line, we obtain

 \beq
\delta m ( T ; h ) = -
2  ( J_{1,1} (T) - J_{2 , 2}   (T) )^2 \Xi ( h ) 
\:\:\:\: , 
\label{pertu.29b}
\eneq
\noindent
with 

 \beq
\Xi ( h ) = - \left( \frac{2}{\pi} \right)^3 \: 
\int_{ - 2 J - H}^{2J - H } \prod_{ i = 1}^3  \left[ \frac{d \epsilon_i}{2 J}
\sqrt{1 - \left( \frac{\epsilon_i + H}{2 J} \right)^2}\right]
\left[ \frac{\epsilon_1}{ ( | \epsilon_1 | + | \epsilon_2 |  ) ( | \epsilon_1 |  + | \epsilon_3 | ) 
( | \epsilon_2 |  + | \epsilon_3 ) |  } \right] 
\:\:\:\: . 
\label{pertu.26}
\eneq
\noindent
In Fig.\ref{plot_delta}
we draw the four functions ${\cal F}_j [ \gamma ]$, $j = 1 , 2 , 3 ,4$, defined in Eqs.(\ref{pertu.28}), 
as functions of $\gamma$. As expected, ${\cal F}_1 [ \gamma ] , {\cal F}_2 [ \gamma ]$ and 
${\cal F}_3 [ \gamma ]$ all converge towards the same value ${\cal F}_*$ as $\gamma \to 1$, that is, 
where the CKL meets the $XX$-line, with ${\cal F}_* \approx 0.21$. Also as expected, one gets 
$\lim_{ \gamma \to 1} {\cal F}_4 [ \gamma ] \approx 0.62 = 3 {\cal F}_*$. This is also consistent with the plots 
of Fig.\ref{plot_h}, where we display $\Xi ( h )$ as a function of $h$ for $-1 \leq h \leq 1$. As expected, 
one obtains $\lim_{ h \to 1} \Xi ( h ) = \Xi_* \approx 0.21$, that is, $\Xi_* \approx {\cal F}_*$, consistent
with the ''collapse'' of Eq.(\ref{pertu.29}) onto Eq.(\ref{pertu.29b}) at the intersection point 
between the CKL1 and the $XX$-line. 

\noindent
\begin{figure}
\includegraphics*[width=.9\linewidth]{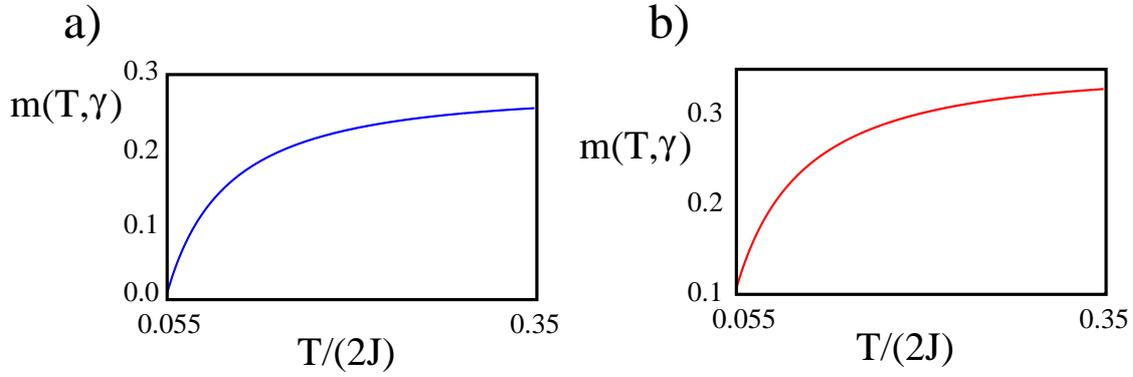}
\caption{Plot of $m ( T , \gamma )$ {\it vs.} $T$ in the
interval $0.055 [(1+\gamma) J] \leq T \leq  [(1+\gamma) J] / \pi$ for 
 $J_{1 ,1}^{(0)} = 0.5 [ J ( 1 + \gamma )]$ and $J^{(0)}_{2 , 2} = \gamma J_{1 , 1}^{(0)}$ and  for 
two paradigmatic values of $\gamma$: 
{\bf a) }  $m ( T , \gamma )$ {\it vs.} $T$ for $0.055 [(1+\gamma) J] \leq T \leq  [(1+\gamma) J] / \pi$ and 
$\gamma = 0.005$;
{\bf b) }  $m ( T , \gamma )$ {\it vs.} $T$ for $0.055 [(1+\gamma) J] \leq T \leq  [(1+\gamma) J] / \pi$ and 
$\gamma = 0.5$. 
In both cases, on lowering $T$,  the reduction in $m ( T , \gamma )$ due to the onset of the Kondo regime 
can be clearly seen.
} \label{running_1}
\end{figure}

\begin{figure}
\includegraphics*[width=.9\linewidth]{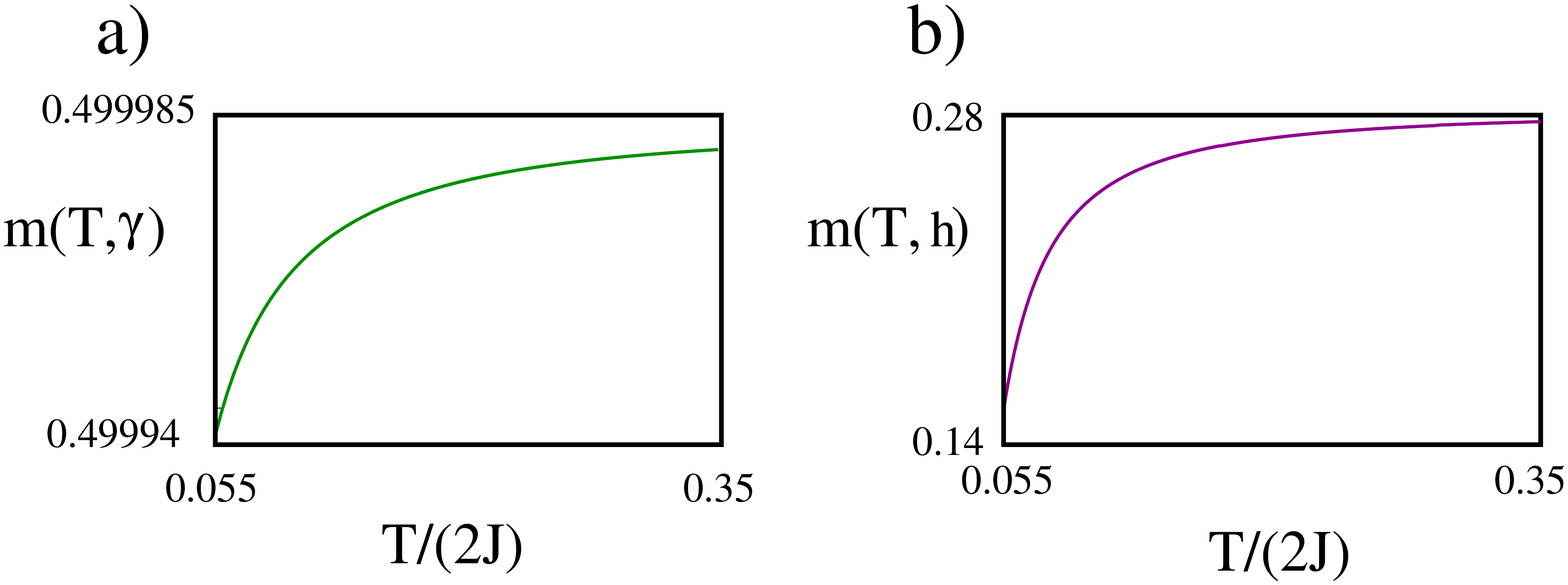}
\caption{{\bf a) } Plot of  $m ( T , \gamma )$ {\it vs.} $T$ in the interval 
$0.055 [(1+\gamma) J] \leq T \leq  [(1+\gamma) J] / \pi$  for 
 $J_{1 , 1}^{(0)} = 0.5 [ J ( 1 + \gamma )]$ and $J^{(0)}_{ 2  , 2} = \gamma J_{1 , 1}^{(0)}$ and  for  
$\gamma = 0.995$. As expected (see text), there is a rather small percentual 
change in $m ( T , \gamma )$ across the interval of values of $T$ that we consider; 
{\bf b)} plot of $m ( T , h )$  {\it vs.} $T$ in the interval 
$0.055 ( 2 J ) \leq T \leq  2 J  / \pi$ for $h = 0.5$ in 
the case of asymmetric couplings $J_1^{(0)}  = 0.5 (2J), J_2^{(0)} = 0.35 (2J)$. In this 
case, the running of $m ( T , h )$ is again apparent. 
} \label{running_2}
\end{figure}

\noindent
Combining Eqs.(\ref{pertu.29},\ref{pertu.29b}) with the numerical solutions of 
Eqs.(\ref{cro.6}), we are ultimately able to recover the RG-flow of $m ( T )$ along 
both the CKL1 at fixed $\gamma$ ($m ( T , \gamma )$), and the $XX$-line 
at fixed $h$ ($m ( T , h )$).   As paradigmatic
cases, in Fig.\ref{running_1}{\it a)} we plot the curve corresponding to 
$m ( T , \gamma = 0.005 )$ (close to the Ising limit) and in Fig.\ref{running_1}{\it b)} 
the curve corresponding to  $m ( T , \gamma = 0.5)$. The interval of values of $T$ 
we chose ends at   $T \sim 2 J ( 1 + \gamma )  / ( 2 \pi )$, consistent 
with a full bandwidth of $ 2 J ( 1 + \gamma) $ and with our choice of temperature units, 
and starts at  $T /(J ( 1 + \gamma) ) \sim 0.055$, where a sensible reduction in 
$m ( T )$ appears as a quite clear evidence 
of the onset of the Kondo regime. To realize both plots in Fig.\ref{running_1}, we have 
set $J_{1 , 1}^{(0)} / ( J ( 1 + \gamma)) = 0.5$ and $J_{2 , 2}^{(0)} = \gamma' J_{1 , 1}^{(0)}$, 
with $\gamma' = \gamma$. A remarkable result is what appears in Fig.\ref{running_2}{\bf a)},
where we plot $m ( T , \gamma = 0.995 )$ as a function of $T$. In fact, due to the 
proximity to the  $XX$-line and to the symmetric situation  $J_{1,1}^{(0)} = J_{2,2}^{(0)}$,
$m ( T ,  \gamma = 0.995 )$ hardly flows, on lowering $T$, as it can be seen from the plot, where, 
within the same interval of values of $T$ as in Fig.\ref{running_1}{\bf a)} and in 
Fig.\ref{running_1}{\bf b)}, $m$ varies by less than one part over $10^4$, consistent 
with the fact that, as we discuss in the following, 
there is no expected flow in $m(T)$ induced by the boundary interaction 
along the $XX$-line in the case of symmetric coupling. For the sake of completeness, 
in Fig.\ref{running_2}{\bf b)} we also report the flow of $m ( T , h  )$ for $h = 0.5$ in 
the case of asymmetric couplings $J_{1 , 1}^{(0)}  = 0.5 (2J), J_{2 , 2}^{(0)} = 0.35 (2J)$. In this 
case, the running of $m ( T , h )$ is again apparent. 
\noindent
\noindent
Before continuing with our discussion, it is now worth focusing onto the 
symmetric case $J_{1,1}^{(0)} = J_{2,2}^{(0)}$, which we left aside at the 
beginning, as a special situation. Indeed, in this case, since the symmetry 
condition among the boundary coupling strengths is preserved along the RG-trajectories, 
that is, since $J_{1,1} ( T ) = J_{2,2} ( T )$ at any $T$, and because of the commutation
relation

 \beq
\left[ \frac{1}{3} \sum_{ \lambda = 1}^3  S_{ 1 , \lambda}^z
, \sum_{ \mu = 1}^3 \left( S_{ 1 , \mu  }^x S_{ 1 , \mu +1 }^x 
+ S_{ 1 , \mu }^y S_{ 1 , \mu +1 }^y \right) \right] = 0
\:\:\:\: ,
\label{pertu.27}
\eneq
\noindent
one obtains that the magnetization is an exactly conserved
quantity of the {\it full} Hamiltonian $H_S=H_{XY} + H_\Delta$ and, 
accordingly, that it is not renormalized by the Kondo interaction. 
This is definitely consistent with the plot drawn in Fig.\ref{running_2}{\bf a)}, 
where practically no flow of $m ( T , \gamma )$ as a function of $T$ appears. 
Since this is clearly a consequence of the enhanced symmetry of the 
whole Hamiltonian along the $XX$-line, the absence of flow in $m ( T )$ can therefore
be related to a recovering of the 4CK-effect along this line in the symmetric case. 

\begin{figure}
\includegraphics*[width=.6\linewidth]{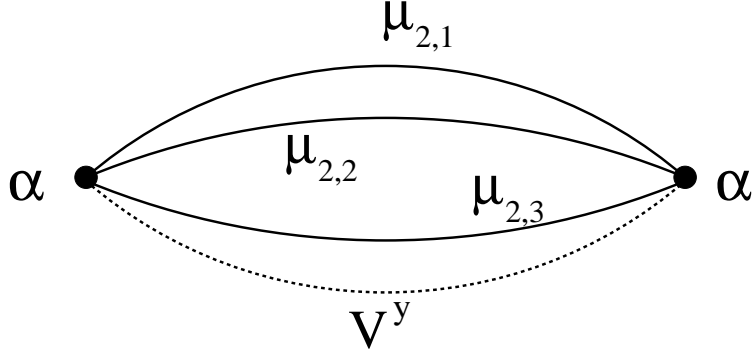}
\caption{Diagrammatic representation of the leading contribution to the 
strong coupling partition function from the effective boundary Hamiltonian 
$H_{\rm Sc; B}^{(2)}$. Specifically, $\alpha = \frac{1}{2} \left( \frac{J_{ 2 , 2} }{J_{ 1 , 1}} 
\right)  [ J ( 1 + \gamma ) + B ]$, the full lines represent the propagation of the 
$\mu$-fermion and the dashed line corresponds to the propagation of 
the $V^y$-operator.
} \label{strong_bubble}
\end{figure}

Based on the evidence arising from the analytical calculation and on plots such as the ones 
drawn in Figs.\ref{running_1},\ref{running_2}, 
we propose to look at the local magnetization $m ( T )$ as  a probe of the 
onset of the Kondo regime. While in the perturbative regime in $J_{1 , 1} , J_{2 , 2}$ the 
decrease in the local   magnetization is mainly due to logarithmic corrections to 
$J_{1 , 1} ( T ) , J_{2 , 2} ( T )$  taking off on lowering $T$, the scaling   law with 
which $m ( T , \gamma )$ or $m ( T , h )$   flow to zero at the strongly-coupled fixed point 
can be eventually inferred from the strong-coupling effective theory. To do so, 
we note that, as we outline  in   \ref{spindensities}, 
at the strongly-coupled fixed point, all the 
operators at site-1 are ''fused'' with the topological spin operators and disappear
from the effective boundary Hamiltonian $H_{\rm Sc}^{(2)}$. In order to compute
$m$ close to the strongly-coupled fixed point we therefore resort to an alternative 
strategy, that is, we add to the Hamiltonian $H_{XY} + H_\Delta$ a ''source term'' $H_B$ for 
the local magnetization, given by 

 \beq
H_B = - \frac{B}{3} \: \sum_{ \lambda = 1}^3 S_{1, \lambda}^z = 
 - \frac{i B}{6} \: \sum_{ \lambda = 1}^3 \mu_{ 1 , \lambda } \mu_{ 2 , \lambda} 
 \:\:\:\: . 
 \label{scmag.1}
 \eneq
 \noindent
We therefore compute the partition function at nonzero $B$, ${\cal Z} [ B ] $, close 
to the strongly-coupled fixed point and eventually calculate $m$ as 

\beq
m = \frac{1}{\beta} \: \frac{\partial \ln {\cal Z} [B] }{\partial B} \Biggr|_{ B  = 0 }
\:\:\:\: . 
\label{smag.2}
\eneq
\noindent
Using Eq.(\ref{scm.7}) for the residual boundary Hamiltonian at the strongly 
coupled fixed point, one may compute $\ln {\cal Z} [B] $ to leading order in 
$H_{\rm Sc ;B}^{(2)}$. The leading nonzero contribution comes from processes 
like the ones sketched in the Feynman diagram we draw in Fig.\ref{strong_bubble}, 
where the vertex $\alpha = \frac{1}{2} \left( \frac{J_{2 , 2}}{J_{1 , 1}} \right) 
[ J ( 1 + \gamma ) + B ]$, the full lines represent the propagation of the 
$\mu$-fermion and the dashed line corresponds to the propagation of 
the $V^y$-operator. The analytical result is given by 

 \beq
\ln {\cal Z} [B]  \approx \ln \tilde{\cal Z}_0 + \frac{1}{8} \left( \frac{J_{2 , 2}}{J_{ 1 ,
1} } \right)^2 
[ J ( 1 + \gamma ) + B ]^2 \: \int_0^\beta \: d \tau_1 \: d \tau_2 
\: [  G_{2,2}  ( \tau_1 - \tau_2 ) ]^3 \: {\cal G}_{\bf V} ( \tau_1 - \tau_2 ) 
\:\:\:\:  ,
\label{smag.3}
\eneq
\noindent
with $\tilde{\cal Z}_0$ being the partition function for the three disconnected wires
with all the terms involving the operators $\mu_{ 1 , \lambda }$ dropped out
of the Hamiltonian $H_{XY}$, ${\cal G}_{\bf V} ( \tau_1 - \tau_2 ) \equiv
\langle {\bf T}_\tau {\bf V}^y ( \tau_1 ) {\bf V}^y ( \tau_2 ) \rangle = {\rm sgn} ( \tau_1 - 
\tau_2 )$,  and $ G_{2 , 2} ( \tau_1 - \tau_2 )$ given in Eq.(\ref{exp.18}).
Without entering the details, we may readily infer the large-$\tau$ limit of $ G_{2,2} ( \tau )$
by only retaining leading contributions, as $\epsilon \to 0$,  in the argument of the corresponding 
integral in Eq.(\ref{exp.18}).
\noindent
\begin{figure}
\includegraphics*[width=.7\linewidth]{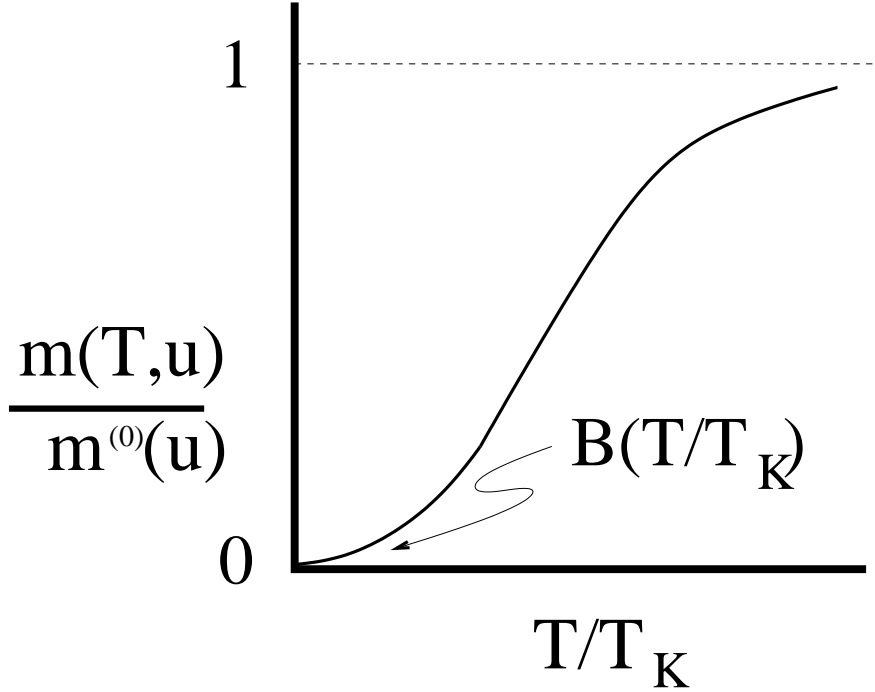}
\caption{Plot of the  magnetization as a function of the temperature.} 
\label{summary}
\end{figure}
\noindent
The corresponding result is 

 \beq
 G_{2,2}  ( \tau ) \approx_{ | \tau | \to \infty} \: \frac{ 2 ( 1 + \delta )^2}{\pi J ( 1 + \gamma ) \delta^3 } \:
\int_0^{ 2 J ( 1 + \gamma ) } \: d \epsilon \: \left( \frac{\epsilon}{J ( 1 + \gamma ) } \right)^2 
e^{  - \epsilon | \tau | }   {\rm Sgn}  ( \tau ) 
\:\:\:\:.
\label{smag.5}
\eneq
\noindent
Using the result in Eq.(\ref{smag.5}), one readily finds that the integral at the 
term $\propto [ J ( 1 + \gamma ) + B ]^2$ at the right-hand side of Eq.(\ref{smag.3}) is independent 
of $\beta$ (that is, independent of $T$) - see the plots in Fig.\ref{magne}. Therefore,
from Eq.(\ref{smag.2}), we are led to conclude that, as $T \to 0$, $m ( T ) = B T / T_K$, with $B$ being a 
numerical constant. Knowing that, consistently with the analysis of the phase diagram reported in
\cite{coleman}, there are no intermediate-coupling fixed
points between the weakly- and the strongly-coupled ones, we may
infer that, on lowering $T$,   $ m ( T , u ) $ (with $u= \gamma$ along the CKL1 and $u= h$ along 
the $XX$-line) starts from $m^{(0)} ( u ) $ and takes a quadratic 
correction in $J_{1 , 1}  , J_{2 , 2}$, which logarithmically increases with $T$ when approaching 
 $T_K$. Eventually, the diagram turns into a linear dependence 
of $m ( T , u )  $ on $T / T_K$, as $ T \to 0$, finally flowing to $0$ at
the strongly-coupled fixed point. The corresponding plot is expected to be quite
alike  to the one reported in Fig.\ref{summary}, an analog of which has been 
presented in Fig.4 of \cite{giu_so_epl_2}, for a junction of three 
quantum Ising chains, with the running coupling $D$ to be identified with $T$. 
Such a scaling diagram is the signature of the onset of 
the Kondo regime. In particular, the linear dependence of $m ( T , u ) $ on 
$ T / T_K$ for $T / T_K \to 0$ is the fingerprint of the 2CK
effect \cite{aflud_1}, which, in this system, takes a peculiar realization, not
requiring any fine-tuning of the couplings between itinerant electrons and
the magnetic impurity \cite{tsve_1}. Thus, we conclude that an experimental measurement of 
$m ( T , u ) $ provides an effective tool to monitor the onset of the 2CK effect.

\begin{figure}
\includegraphics*[width=.8\linewidth]{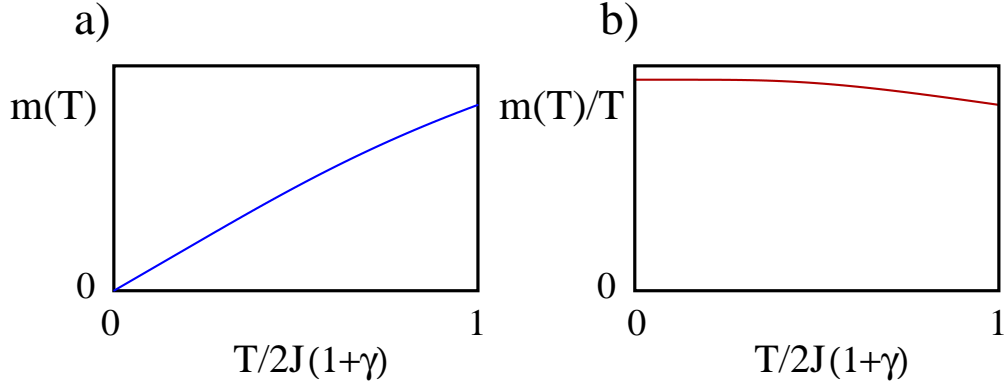}
\caption{ {\bf a)} Plot of $m(T)$ (arbitrary units) as derived from Eqs.(\ref{smag.2},\ref{smag.3}) across
a range of $T$ from 0 to the full bandwidth: as $T \to 0$ (relevant part of 
the plot) there is an apparent linear dependence on $T$; 
{\bf b)} The same plot as in panel {\bf a)}, but multiplied by $\beta (=1/T)$: the 
flow towards a constant value as $T \to 0$ enforces the linearity of the plot in 
panel {\bf a)} as $T \to 0$, but multiplied by $\beta (=1/T)$: the 
flow towards a constant value as $T \to 0$ enforces the linearity of the plot in 
panel {\bf a)} as $T \to 0$.} 
\label{magne}
\end{figure}
\noindent

\section{Conclusions}
\label{sec:concl}

In this paper we have considered a $Y$-junction of anisotropic $XY$ spin chains in a magnetic field,
with the chains coupled to each other at a central region determined by the interaction
between their initial spins. In general, spin chains offer the possibility of simulating magnetic impurities 
and to engineer tunable low-energy effective multi-channel Kondo Hamiltonians, in
which the symmetry between channels is enforced by RG-flow to low-temperatures/energies
\cite{tsve_1}, differently with what happens with multi-channel Kondo effect in 
semiconducting devices, such as, for instance, quantum dots, where   
without fine tuning the parameters, an anisotropy between the two channel emerges, 
making the coupling to a single channel dominant  \cite{giuliano02,oreg03,oreg07}.
In addition, the anisotropic $XY$ model in a magnetic field allows for 
continuously tuning the bulk parameters keeping the excitation spectrum
gapless. Specifically, we showed that, on continuously tuning the bulk
parameters of our junction, it is possible to move from a junction of 
three $XX$-spin chains in an applied magnetic field, to a $Y$-junction 
of three critical quantum Ising chains, and, in particular, evidenced 
how this corresponds to an evolution from a four-channel Kondo (4CK) to a two-channel Kondo 
(2CK) effective 
Hamiltonian. Moreover, we highlighted that transition from 4CK- to 2CK-effect takes place 
in a discontinuous way at the merging points between the $XX$-line and the critical 
Kitaev lines of the $XY$-chains. 

The scenario emerging from our analysis implies that, in the 
case of symmetric boundary couplings ($J_1 = J_2 $), the 
$XX$-line corresponds to a 4CK-critical line, separating two
2CK-phases, towards either one of which the system flows, once 
the symmetry in the boundary couplings is broken ($J_1 \neq J_2 $). 
Out of the two 2CK-phases separated by the symmetric $XX$-line, one
is continuously connected with the 2CK-phase describing the junction
along the critical Kitaev lines. Thus, our results point out to the possibility 
that, at symmetric boundary couplings, the $XX$-line works as a critical line, 
separating two non-Fermi liquid phases. We believe that this result is 
important, as it implies the realization, in our junction, of a remarkable quantum phase 
transition between two non-Fermi liquid phases. In our view, it 
would be quite interesting to characterize such a quantum phase transition 
in analogy to what has been done for a quantum dot device in \cite{pustilnik04}, where the 
2CK-point has been regarded as the quantum critical point in parameter space, separating two
one-channel Kondo Fermi-liquid phases. Nevertheless,  
a careful characterization of such a quantum phase transition lies  
beyond the scope of this work, and we 
plan to leave it aside, for a future investigation.

We finally mention that in the continuous limit both at 
the $XX$ point and at the critical Ising point the junction boundary term can be exactly treated  
by Bethe ansatz \cite{tsve_3,tsve_4,Buccheri2015}. This suggests a further interesting 
development of our research toward studying whether the exact solvability can be extended 
to the lines of gapless spectrum we discussed in this paper.

\section*{Acknowledgments}
The authors very gratefully acknowledge useful discussions with H. Babujian, 
F. Buccheri, N. Cramp\`e, L. Dell'Anna,   F. Franchini, V. E. Korepin, and A. A. Nersesyan. 
They also thank M. Fagotti for a very useful and enlightening 
correspondence about the solutions of the $XY$-chain with OBC.
 
{\it Note Added:} During the final stage of the work for this paper, a very interesting 
paper by A. A. Nersesyan and M. M\"uller on 
the response of classical impurities in quantum Ising chains appeared on the arXiv \cite{muller16}.

\appendix

\section{Spin-isospin representation of the two-channel Kondo Hamiltonian}
\label{spindensities}

In this Appendix we discuss a pertinently adapted version of  
the spin-isospin representation of the spin-$1/2$ two-channel
Kondo Hamiltonian introduced in \cite{coleman}, which we use 
as a guideline to discuss  the effective Kondo interaction emerging 
in our junction. As a starting point, on each chain one trades the 
$3 \ell$ $c_{ j , \lambda }$-operators (with $j = 1 , \ldots , \ell$) 
for $6 \ell$ real fermion  operators $\mu_{ j , \lambda}$ ($j = 1 ,\ldots , 2 \ell$), 
related to each other via

 \beq
c_{ j , \lambda } = \frac{1}{2} \left( \mu_{ 2j - 1 , \lambda } + i \mu_{ 2 j , \lambda }
\right)\:\:\:\: . 
\label{relation_JW}
\eneq
\noindent 
The next step is to introduce an additional set of 
  spinful lattice Dirac fermions $ \{ d_{ j , \sigma } , d_{ j ,\sigma}^\dagger \}$,
  with $\sigma= \uparrow, \downarrow$ and $j = 1 , \ldots , 2 \ell$. To do so, 
following \cite{coleman} 
we introduce a fourth  lattice real fermion 
$ \{ \mu_{ j , 0 } \}$ ($j = 1 , \ldots , 2 \ell$), which, as we will 
eventually check, has to decouple   from the Kondo boundary interaction and
from any relevant physical observable. Therefore, we set

 \begin{eqnarray}
 d_{ j , \uparrow} &=& \frac{1}{2} \left( \mu_{ j , 1} + i \mu_{ j , 2 } \right) \nonumber \\
 d_{ j , \downarrow} &=& \frac{1}{2} \left( \mu_{ j , 3} + i \mu_{ j , 0 } \right)
  \:\:\:\: , 
  \label{sd.1}
\end{eqnarray}
\noindent
with $j = 1 , \ldots , 2 \ell$ [notice that Eqs.(\ref{sd.1}), though local in the spin index, 
are non-local in the chain index]. In terms of the $d$-fields one may therefore 
rewrite $H_{XY}$ as 

 \begin{eqnarray}
 H_{XY} &=&  - \frac{i J }{2} \: 
\sum_{ j = 1}^{ 2 \ell - 1 } \mu_{ j, 0} \mu_{ j + 1 ,0} 
-  \frac{i \gamma J}{4} \: 
\sum_{ j = 1}^{ 2 \ell - 3 } \delta_j \: 
\mu_{ j , 0 } \left( \mu_{ j + 1 , 0 } - \mu_{  j + 3 ,  0 } \right)  
\nonumber \\ 
&+& i J \: \sum_\sigma \: \sum_{ j = 1}^{ 2 \ell - 1 } \: 
\left(  d_{ j , \sigma}^\dagger d_{  j + 1  , \sigma } - d_{ j + 1  , \sigma}^\dagger 
d_{ j , \sigma} \right) \nonumber \\
&+&
\frac{i \gamma J}{2} \: \sum_\sigma \: \sum_{ j = 1}^{ 2 \ell - 1} 
 \:  \delta_j \, \left(  d_{ j , \sigma}^\dagger  d_{ j + 1 , \sigma } -
 d_{ j +1 , \sigma}^\dagger d_{ j , \sigma} \right) \nonumber \\
 &-& \frac{i \gamma J}{2} \: \sum_\sigma \: \sum_{ j = 1}^{ 2 \ell - 3} 
  \:  \delta_j \, \left(  d_{ j , \sigma}^\dagger  d_{  j + 3  , \sigma } - 
   d_{ j + 3  , \sigma}^\dagger d_{ j , \sigma} \right)  
\:\:\:\: , 
\label{sd.2}
\end{eqnarray}
\noindent
where $\delta_j \equiv 1  - ( - 1 )^j$. In order to express $H_\Delta$ in terms of the $d$-fermions, 
one defines two commuting lattice vector density operators, a lattice spin density 
$\vec{\sigma}_j$ and an isospin density $\vec{\tau}_j$, given by 

 \beq
\vec{\sigma}_j = \left( \begin{array}{c} \sigma_j^1 \\ \sigma_j^2 \\ 
                         \sigma_j^3
                        \end{array} \right) = 
\frac{1}{2} \left( \begin{array}{c}
d_{ j , \uparrow}^\dagger d_{ j , \downarrow} + d_{ j , \downarrow}^\dagger 
d_{ j , \uparrow} \\ -i ( d_{ j , \uparrow}^\dagger d_{ j , \downarrow} - 
d_{ j , \downarrow}^\dagger d_{ j , \uparrow}) \\
d_{ j , \uparrow}^\dagger d_{ j , \uparrow} - d_{ j , \downarrow}^\dagger 
d_{ j , \downarrow} 
                   \end{array} \right) 
                   \;\;\; , \;\;
\vec{\tau}_j = \left( \begin{array}{c} \tau_j^1 \\ \tau_j^2 \\ 
                         \tau_j^3
                        \end{array} \right) = 
\frac{1}{2} \left( \begin{array}{c}
d_{ j , \uparrow}^\dagger d_{ j , \downarrow}^\dagger + d_{ j , \downarrow} 
d_{ j , \uparrow} \\ -i ( d_{ j , \uparrow}^\dagger d_{ j , \downarrow}^\dagger - 
d_{ j , \downarrow}  d_{ j , \uparrow} ) \\
d_{ j , \uparrow}^\dagger d_{ j , \uparrow} + d_{ j , \downarrow}^\dagger 
d_{ j , \downarrow} -1
                   \end{array} \right)
                   \;\;\;\; . 
                   \label{sd.3}
                        \eneq
                        \noindent
In terms of the operators in Eq.(\ref{sd.3}), one therefore obtains  
 
 \beq
 \vec{\Sigma}_j = \vec{\sigma}_{ 2 j - 1 } + \vec{\tau}_{ 2 j - 1 } \;\;\; , \;\;
 \vec{\Upsilon}_j = \vec{\sigma}_{ 2 j } + \vec{\tau}_{ 2 j  }
 \:\:\:\: , 
 \label{sd.4}
 \eneq
 \noindent
from which one obtains
 
 \beq
H_\Delta = 2 J_\Delta \left[ \left( \vec{\sigma}_1 + \vec{\tau}_1 \right) + \gamma' \left( 
\vec{\sigma}_2 + \vec{\tau}_2 \right) \right] \cdot \vec{\cal R} 
\equiv 2  \left[ J_1 \left( \vec{\sigma}_1 + \vec{\tau}_1 \right) + J_2 
\left( \vec{\sigma}_2 + \vec{\tau}_2 \right) \right] \cdot \vec{\cal R} 
\:\:\:\: . 
\label{sd.5}
\eneq
\noindent
From Eq.(\ref{sd.5}) one clearly sees that the $_0$-lattice fermions decouple from $H_\Delta$, 
as it should be for the mapping procedure to be consistent. Also, 
from Eq.(\ref{sd.5}) it appears that $H_\Delta$ describes
two pairs of  independent spin-$1/2$ lattice density operators, antiferromagnetically
coupled to the isolated impurity $\vec{\cal R}$, with coupling strengths 
$J_1 = J_\Delta$ and $J_2 = \gamma' J_\Delta$. 

The description in terms of the $d$-spin operators is also effective in 
working out the ''residual'' boundary interaction at the Kondo fixed point, 
which we used in the main text to discuss the stability of the 
putative strongly-coupled fixed point in the various regimes. An important 
preliminary remark is that, on employing the realization of the 
two-channel spin-$1/2$ Kondo model in terms of spin- and isospin-density operators
coupled to a spin-$1/2$ impurity, the isotropic two-channel overscreened Kondo fixed point, which, 
in the ''standard'' realization  is known to take place at a finite coupling
\cite{blandin,aflud_1,aflud_3}, is pushed towards an effective  
strongly-coupled fixed point \cite{coleman}. As discussed in the main text, from the expression 
of $H_\Delta$ we expect that, when the bare coupling $J_1$ is $> (<) J_2$,
the strongly-coupled fixed point of the $Y$-junction 
corresponds the 2CK-fixed point where the two channels coupled to 
via $J_1$ ($J_2$) are strongly-coupled to $\vec{\cal R}$. Assuming for the 
sake the discussion that $J_1$ flows to strong coupling first, 
in the strongly-coupled limit the groundstate of the system is 
recovered by minimizing the boundary interaction $\propto J_1$. This can be done by noticing that 
states carrying at site-$j$ spin-$1/2$ associated to the $\vec{\sigma}_j$-operators 
carry spin $0$ associated to the $\vec{\tau}_j$-operator, and vice versa.  
As a result, at strong coupling, the low-energy manifold
consists of the spin singlets $ | \Sigma \rangle_{ 1 , 2}$, defined as 

 \begin{eqnarray}
 | \Sigma \rangle_1 &=& \frac{1}{\sqrt{2}} \{ d_{1 , \uparrow}^\dagger  | \Downarrow \rangle 
 -  d_{1 , \downarrow}^\dagger | \Uparrow \rangle \} 
 \nonumber \\
  | \Sigma \rangle_2 &=& \frac{1}{\sqrt{2}} \{  d_{1 , \uparrow}^\dagger d_{1 , \downarrow}^\dagger| \Downarrow \rangle 
 - | \Uparrow \rangle \} 
 \;\;\;\; . 
 \label{sd.6}
\end{eqnarray}
\noindent
with  $| \Uparrow \rangle , | \Downarrow \rangle$ being the spin $\pm 1/2$ eigenstates of ${\cal R}^3$, 
with $\vec{\cal R} \equiv \left( {\cal R}^1 \: , \: {\cal R}^2 \: , \: {\cal R}^3 \right)^T$ 
given by Eq.(\ref{xy.7}). 
At large, but finite, coupling $J_1$, the system can undergo virtual transitions 
from either singlet in Eqs.(\ref{sd.6}) to excited states and back to 
the singlets. These are induced by the ''residual'' boundary Hamiltonian
$H'$, given by 

 \beq
H' = i J \sum_\sigma \left( d_{ 1 , \sigma}^\dagger O_\sigma - O_\sigma^\dagger d_{ 1 , \sigma } \right) 
+ 2 J_2 \vec{\cal R} \cdot \left( \vec{\sigma}_2 + \vec{\tau}_2 \right) \equiv 
H^{'}_1 + H_2^{'} 
\;\;\;\; , 
\label{sd.7}
\eneq
\noindent
with $O_\sigma \equiv \left( 1 + \gamma \right) d_{ 2 , \sigma } - \gamma d_{ 4 , \sigma }$. The summation 
over virtual transitions can be performed within a systematic Schrieffer-Wolff  procedure, which
yields to the effective boundary Hamiltonian $H_{\rm Sc}^{(2)}$ whose matrix 
elements between the singlets $ | \Sigma \rangle_{j}$ and $ | \Sigma \rangle_{ j ' } $ are given by

 \beq
[ H_{\rm Sc}^{(2)} ]_{ j , j'} \approx  - \frac{1}{J_1 } \: \sum_{ X} \: _j \langle \Sigma | H' | X \rangle 
\langle X |  H' | \Sigma \rangle_{ j'} 
\:\:\:\: , 
\label{sd.8}
\eneq
\noindent
with the sum over the intermediate states $X$ being carried over the (locally) excited triplet states given by 
 
 \begin{eqnarray}
 | T_1 \rangle_1 &=& d_{1 , \uparrow}^\dagger  | \Uparrow \rangle  \; , 
\;  | T_0 \rangle_1  =  \frac{1}{\sqrt{2}} \left( d_{1 , \uparrow}^\dagger | \Downarrow \rangle + 
 d_{1 , \downarrow}^\dagger | \Uparrow \rangle \right)  \; , 
\;   | T_{-1} \rangle_1  = d_{1 , \downarrow}^\dagger | \Downarrow \rangle  \nonumber \\
| T_1 \rangle_2 &=& d_{1 , \uparrow}^\dagger  d_{1 , \downarrow}^\dagger    | \Uparrow \rangle \; , \; 
  | T_0 \rangle_2 = \frac{1}{\sqrt{2}} \left( d_{1 , \uparrow}^\dagger d_{1 , \downarrow}^\dagger    | \Downarrow \rangle + 
    | \Uparrow \rangle \right)
\; , \; 
  | T_{ -1} \rangle_2  =   | \Downarrow \rangle
  \:\:\:\: . 
  \label{sd.9}
\end{eqnarray}
\noindent
On explicitly computing the sum, one obtains the final result which, once expressed back in terms
of the $\mu$-fermions, reads 
 
 \begin{eqnarray}
[ H_{\rm Sc}^{(2)} ]_{ 1 , 1} &=& [ H_{\rm Sc}^{(2)} ]_{ 2 , 2 } = - \frac{ J_2^2 }{ 2 J_1 } \left( \frac{3}{2} + 
 2 \vec{\sigma}_2 \cdot \vec{\tau}_2 \right) - \frac{3 J^2}{4 J_1 } = - \frac{3 J_2^2}{4 J_1 }
 - \frac{3 J^2}{4 J_1 } 
 \nonumber \\
[ H_{\rm Sc}^{(2)} ]_{ 1 , 2}&=& - [ H_{\rm Sc}^{(2)} ]_{ 2 , 1}=  \frac{J J_2}{2 J_1} 
  \left[ (1 + \gamma ) \mu_{ 2 , 1 } \mu_{ 2 , 2 } \mu_{ 2 , 3} - 
  \gamma \left( \mu_{ 2,1} \mu_{ 4 , 2} \mu_{ 2 , 3 } + \mu_{2,1} \mu_{ 2 , 2 } 
  \mu_{ 4 , 3 } + \mu_{ 4 , 1 } \mu_{ 2 , 2 } \mu_{ 2 , 3 } \right)  \right] \nonumber \\
  \:\:\:\: . 
  \label{sd.10}
\end{eqnarray}
\noindent
As it appears from Eqs.(\ref{sd.10}), up to a constant, we 
can write the effective strong-coupling Hamiltonian as 

 \beq
H_{\rm Sc}^{(2)}  =   i \frac{J J_2}{2 J_1} {\bf V}^y \left[ (1 + \gamma ) \mu_{ 2 , 1 } \mu_{ 2 , 2 } \mu_{ 2 , 3} - 
  \gamma \left( \mu_{ 2,1} \mu_{ 4 , 2} \mu_{ 2 , 3 } + \mu_{2,1} \mu_{ 2 , 2 } 
  \mu_{ 4 , 3 } + \mu_{ 4 , 1 } \mu_{ 2 , 2 } \mu_{ 2 , 3 } \right)  \right] 
  \;\;\;\; . 
  \label{sd.11}
\eneq
\noindent
In Eq.(\ref{sd.11}) it is ${\bf V}^y \equiv  \mu_{ 1 , 1 } \mu_{ 1 , 2} \mu_{ 1 , 3}$, with ${\bf V}^y$ is the 
operator exchanging the two singlets: 
  
  \beq
{\bf V}^y | \Sigma \rangle_1 = -i | \Sigma \rangle_2 \;\;\; , \;\;
{\bf V}^y | \Sigma \rangle_2 = i| \Sigma \rangle_1
\:\:\:\: . 
\label{sd.12}
\eneq
\noindent
An important point to stress is that $H_{\rm Sc}$ in Eq.(\ref{sd.11}) is second-order in
$H'$ and, specifically, it is $\propto J J_2$. When $J_2 = 0$ 
(meaning that $\gamma'$ is set to $0$ from the very beginning), $H_{\rm Sc}^{(2)} $
is zero, besides a trivial constant shift of the energies. In this case, as it happens in the 
analysis of the ''standard'' 2CK-problem in \cite{coleman}, the residual boundary interaction at 
strong coupling is recovered to third-order in $H'$. When $J_2 = 0$, extending the SW procedure to 
third-order in $H'$ following, for instance, the approach developed in \cite{giutag}, one obtains 
the effective Hamiltonian $H_{\rm Sc}^{(3)}$ given by 
 
 \beq
H_{\rm Sc}^{(3)} =  i \frac{3 J^3}{4 J_1^2} {\bf V}^y 
[ ( 1 + \gamma ) \mu_{ 2 , 1 } - \gamma \mu_{ 4 , 1 } ] 
  [ ( 1 + \gamma ) \mu_{ 2 , 2 } - \gamma \mu_{ 4 , 2 } ]  
  [ ( 1 + \gamma ) \mu_{ 2 , 3 } - \gamma \mu_{ 4 , 3 } ] 
  \;\;\;\; . 
  \label{sd.13}
  \eneq
  \noindent
The operators in Eqs.(\ref{sd.11},\ref{sd.13}) have been used in the main text
when discussing the stability of the strongly-coupled fixed point in the various 
regions of the system's parameters. An important conclusion 
from Eqs.(\ref{sd.11},\ref{sd.13}) 
is that both $H_{\rm Sc}^{(2)}$ and $H_{\rm Sc}^{(3)}$ correspond to 
irrelevant operators, as they are both obtained as a product of three fermionic fields, with 
resulting scaling dimension $d_{\rm Sc} = \frac{3}{2} > 1$. Since they are by construction 
the leading boundary perturbation at the strongly-coupled fixed point, this implies the
stability of the two-channel state, consistently with the results of 
\cite{tsve_1}.

To conclude this Appendix, we now review how the result in Eq.(\ref{sd.11}) for $H_{\rm Sc}^{(2)}$ is 
modified by adding to the total Hamiltonian a source term $H_B \equiv - \frac{i B}{6} \: \sum_{ \lambda = 1}^3 \mu_{ 1 , \lambda}
\mu_{2 , \lambda}$ for the local magnetization. To begin with, we rewrite $H_B$ in terms of 
the $d$-fermions as 

 \beq
H_B = \frac{iB}{6} \left[ 2 \left( d_{ 2 , \uparrow}^\dagger d_{ 1 , \uparrow} + d_{ 2 , \uparrow} d_{1, \uparrow}^\dagger \right)
+ \left( d_{ 2 , \downarrow} + d_{ 2 , \downarrow}^\dagger \right) \left( d_{1 , \downarrow} + d_{ 1 , \downarrow}^\dagger \right) 
\right]
\:\:\:\: . 
\label{sd.14}
\eneq
\noindent
Letting $H_B$ act onto the groundstates $ | \Sigma \rangle_1 , | \Sigma \rangle_2$, one obtains 
the following nonzero matrix elements at the strongly-coupled fixed point

 \begin{eqnarray}
 ~_2 \langle T , 1 | H_B | \Sigma \rangle_1 &=&  ~_1 \langle T , 1 | H_B | \Sigma \rangle_2 
 = -  \frac{i B \sqrt{2}}{6 } d_{ 2 , \uparrow} \nonumber \\
  ~_2 \langle T , - 1 | H_B | \Sigma \rangle_1 &=&   ~_1 \langle T , - 1 | H_B | \Sigma \rangle_2 = 
  \frac{i B \sqrt{2}}{6 } d_{ 2 , \uparrow}^\dagger \nonumber \\
      ~_2 \langle T , 0 | H_B | \Sigma \rangle_1 &=&     ~_1 \langle T , 0 | H_B | \Sigma \rangle_2
      = - \frac{i B}{6 }  ( d_{ 2 , \downarrow} + d_{ 2 , \downarrow}^\dagger )
      \:\:\:\: , 
      \label{scm.5}
\end{eqnarray}
\noindent
all the other matrix elements being equal to $0$. Using Eqs.(\ref{scm.5}) 
it is now straightforward to repeat the SW procedure to recover the effective 
boundary Hamiltonian at the strongly-coupled fixed point at nonzero $B$, $H_{\rm Sp ; B}^{(2)}$.
The final result is 

 \begin{eqnarray}
H_{\rm Sc ; B}^{(2)}  &=&   i \frac{J J_2}{2 J_1} {\bf V}^y \{  (1 + \gamma ) \mu_{ 2 , 1 } \mu_{ 2 , 2 } \mu_{ 2 , 3} - 
  \gamma [ \mu_{ 2,1} \mu_{ 4 , 2} \mu_{ 2 , 3 } + \mu_{2,1} \mu_{ 2 , 2 } 
  \mu_{ 4 , 3 } + \mu_{ 4 , 1 } \mu_{ 2 , 2 } \mu_{ 2 , 3 } ]  \} \nonumber \\
&-& \frac{J B}{6 J_1} \: \{ O_\uparrow^\dagger d_{ 2 , \uparrow } + 
O_\uparrow d_{ 2 , \uparrow}^\dagger + ( O_\downarrow + O_\downarrow^\dagger ) 
\} {\bf V}^z + \frac{J_2 B}{2 J_1 } \: i \mu_{2,1} \mu_{2,2} \mu_{2,3} {\bf V}^y
\:\:\:\: , 
\label{scm.7}
\end{eqnarray}
\noindent
with ${\bf V}^y$ defined as in Eq.(\ref{cro.a1}) and, by analogy, ${\bf V}^z$ defined so
that ${\bf V}^z | \Sigma \rangle_1 = | \Sigma \rangle_1$, ${\bf V}^z | \Sigma \rangle_2 = 
- | \Sigma \rangle_2$. In the main text, we use the result in Eq.(\ref{scm.7}) to compute the 
local magnetization close to the strongly-coupled fixed point. 
 
\section{Explicit solution of the Kitaev chain with open boundary conditions}
\label{solution_kitaev}
          
In Section \ref{mapping_0} we employ the JW transformation to map a junction of three
quantum $XY$-spin chains onto a junction of spinless fermionic Kitaev chains. Due 
to the presence of the boundary at $j=1$, where the chains interact with each other via 
the boundary Kondo interaction $H_\Delta$, in order to perturbatively account for 
$H_\Delta$ one has to use the disconnected chain limit as reference unperturbed limit. Therefore, 
one has to explicitly determine the energy eigenmodes and eigenfunctions of a single 
Kitaev model obeying open boundary conditions (OBC). At variance with the straightforward 
derivation in the case of periodic boundary conditions \cite{kitaev}, as far as we know, 
only very recently the case of OBC for a lattice chain 
has  been explicitly discussed in detailed way 
in the literature (see the very recent works \cite{maurizio} for the lattice 
model and \cite{muller16} in  the continuum limit). 
For this reason we devote this Appendix to review the procedure of solving the 
Kitaev Hamiltonian corresponding to a single $XY$-chain with OBC. While we 
restrict our derivation to the large-$\ell$ limit, we eventually argue that 
our results are consistent with the ones of \cite{maurizio}, taken in the 
appropriate range of values of the parameters. 

By means of the  JW-transformation, the Hamiltonian for a single $XY$-chain
is mapped onto the fermionic Hamiltonian

  \beq
H_{F} = - \frac{J ( 1 + \gamma ) }{2} \:  
\sum_{ j = 1}^{ \ell - 1} \left( c_{ j  }^\dagger c_{ j + 1  } + 
c_{ j + 1  }^\dagger c_{ j   } \right) 
+  \frac{J ( 1 -  \gamma ) }{2}  
\sum_{ j = 1}^{ \ell - 1} \left( c_{ j  } c_{ j + 1  } + 
c_{ j + 1  }^\dagger c_{ j   }^\dagger \right)
+ H \: 
\sum_{ j = 1}^{ \ell } c_{ j }^\dagger c_{ j } 
\:\:\:\: . 
\label{xy.5a}
\eneq
\noindent
$H_F$ in Eq.(\ref{xy.5a}) is the Kitaev Hamiltonian as presented in \cite{kitaev}
to describe a one-dimensional $p$-wave superconductor, with single-fermion normal 
hopping amplitude  $ w = \frac{J ( 1 + \gamma ) }{2}$, superconducting 
gap $ \Delta  = \frac{J ( 1 - \gamma ) }{2} $, and chemical potential $ \mu = - H$.
 A generic energy eigenmode of $H_F$, $\Gamma_E$, is written as 

 \beq
\Gamma_E = \sum_{ j = 1}^\ell \left( u^E_j c_j + v^E_j c_j^\dagger \right)
\:\:\:\: , 
\label{exp.2}
\eneq
\noindent
with the wavefunctions $u^E_j , v^E_j$ obeying the  Bogoliubov-de Gennes (BdG)-equations for 
a $p$-wave superconductor, given by

 \begin{eqnarray}
 E u^E_j &=& - \frac{J ( 1 + \gamma ) }{2} \left( u^E_{ j + 1 } + u^E_{ j - 1 } \right) + 
 H u^E_j - \frac{J ( 1 - \gamma) }{2} 
 \left( v^E_{ j + 1 } - v^E_{ j - 1  } \right)  \nonumber \\
  E v^E_j &=&   \frac{J ( 1 + \gamma ) }{2} \left( v^E_{ j + 1 } + v^E_{ j - 1 } \right) - 
  H  v^E_j + \frac{J ( 1 - \gamma) }{2}   \left( u^E_{ j + 1 } - u^E_{ j - 1  } \right)
 \:\:\:\: , 
 \label{exp.3}
\end{eqnarray}
\noindent
which are solved by functions of the form  

 \beq
\left( \begin{array}{c}
        u^E_j \\ v^E_j 
       \end{array} \right) \equiv 
 \left( \begin{array}{c}
        u^E  \\ v^E 
       \end{array} \right) e^{ i k j }
       \:\:\:\: , 
       \label{exp.4}
       \eneq
       \noindent
with 

 \begin{eqnarray}
 E u^E &=& - \left[ J ( 1 + \gamma)  \cos{k} - H \right] u^E -  i J ( 1 - \gamma)    \sin{k}  v^E \nonumber \\
 E v^E &=& i J ( 1 - \gamma)  \sin{k}  u^E + \left[ J ( 1 + \gamma)  \cos{k} - H \right] v^E 
 \:\:\:\: . 
 \label{exp.5}
\end{eqnarray}
\noindent
From Eq.(\ref{exp.5}), one readily derives the  dispersion 
relation 

\beq
E_k = \pm \sqrt{ [ J ( 1 + \gamma)  \cos{k} - H ]^2 + J^2  ( 1 - \gamma )^2 \sin^2{k}  }
\:\:\:\: . 
\label{exp.6}
\eneq
\noindent
Before discussing the boundary conditions in $j=1$ and $j = \ell$, we note that, as it appears from Eq.(\ref{exp.6}), 
in general the relation dispersion is gapped, with a minimum energy gap 
$\Delta_w = {\rm min } \{ \Delta_1 , \Delta_2 \}$, where  
$\Delta_1 = J ( 1 - \gamma )   \sqrt{ 1 -  \frac{  H^2}{   \gamma J^2 } } $, 
with the minimum reached at $\cos{k} =  \frac{ ( 1 + \gamma ) H }{2 \gamma J }$, and  
$\Delta_2 = | J ( 1 + \gamma )  - H |$, with the minimum reached at $\cos{k} =  1$. 
For $\gamma = 1$, the dispersion relation in Eq.(\ref{exp.6})   
yields a gapless spectrum as long as
$ - 2 J  \leq H \leq 2 J$. This is the gapless $XX$-line, which, through JW-transformation,
maps onto a free-fermion chain with OBC. As stated above, here we focus onto the 
regime with $H\geq 0$. In this case, for $0 \leq \gamma < 1$, a 
gapless line appears for $H =  J ( 1 + \gamma ) $, with the gap closing at 
$\cos{k} =  1$. This is the critical Kitaev line CKL, 
marking the transition between the topological and the nontopological phase of the 
Kitaev Hamiltonian in Eq.(\ref{xy.5a}). The union of this line with the $XX$-line 
defines  the region in parameter space characterized by a gapless JW-fermion spectrum. 

To discuss the solutions of the BdG equations along the CKL1, 
let us   assume  $H =  J ( 1 + \gamma )  $ and $0 \leq \gamma < 1$. Setting 
$E_k = \pm \epsilon_k$, Eq.(\ref{exp.6}) implies 

 \beq
\epsilon_k = 2 J  \left| \sin \left( \frac{k}{2} \right) \right| 
\: \sqrt{ (1 + \gamma )^2  \sin^2 \left( \frac{k}{2} \right)  + (1 - \gamma )^2
\cos^2 \left( \frac{k}{2} \right) }
\:\:\:\:,
\label{exp.7}
\eneq
\noindent
with the gap closing  at $k=0$ (mod $2 \pi$). To impose the appropriate boundary 
conditions on the open chain, we note that, for generic values of the system 
parameters, forcing the wavefunction in Eqs.(\ref{exp.2}) to
solve Eqs.(\ref{exp.3}), necessarily implies $u^E_{ j = 0 } = v^E_{ j = 0 } = 0$, as 
well as $u^E_{ j = \ell } = v^E_{ j = \ell } = 0$. On assuming $k$ to be real, it is 
easy to check that the only degeneracy in the dispersion relation in Eq.(\ref{exp.6}) is 
for $k \to - k$. This degeneracy 
is however not sufficient to construct solutions satisfying the boundary conditions above, 
which makes it necessary to consider also solutions with complex values of 
the momentum, provided they are normalizable. In fact, at a given value
of $\epsilon_k = \epsilon$, there are two corresponding solutions with 
real momentum $\pm k$ such that 

 \beq
\cos k = \frac{(1+\gamma)^2}{ 4 \gamma } -  \frac{(1 - \gamma)^2}{ 4 \gamma }
\sqrt{1 + \frac{ 4 \gamma  \epsilon^2 }{ J^2 ( 1 - \gamma)^4}}
\:\:\:\: ,
\label{exp.6_bis}
\eneq
\noindent
as well as two solutions with purely imaginary momentum $\pm i q$, such 
that

 \beq
 \cosh q= \frac{(1+\gamma)^2}{ 4 \gamma } +  \frac{(1 - \gamma)^2}{ 4 \gamma }
\sqrt{1 + \frac{ 4 \gamma  \epsilon^2 }{ J^2 ( 1 - \gamma)^4}}
\:\:\:\: .
\label{exp.8_cosh}
\eneq
\noindent
Positive-energy, real-$k$ solutions are constructed by defining   $\alpha_k$ so that 

 \beq
\cos{\alpha_k} = \frac{ ( 1 + \gamma )    \sin \left( \frac{k}{2} \right)   }{
\sqrt{ ( 1 + \gamma)^2 \sin^2  \left( \frac{k}{2} \right)  + 
( 1 - \gamma)^2 \cos^2  \left( \frac{k}{2} \right)}}
\;\;\; ; \;\;
\sin{\alpha_k} =    \frac{( 1 - \gamma )  \cos \left( \frac{k}{2} \right)  }{
\sqrt{ ( 1 + \gamma )^2  \sin^2  \left( \frac{k}{2} \right)  + 
( 1 - \gamma)^2 \cos^2  \left( \frac{k}{2} \right)}}
\:\:\:\:.
\label{exp.9}
\eneq
\noindent
The two independent propagating solutions with real $k$ are therefore given by

\beq
\left( \begin{array}{c}
        u_j^E \\ v^E_j 
       \end{array} \right)_{ \pm k } = 
       \left( \begin{array}{c}
               \cos{\left( \frac{\alpha_k}{2} \right)} \\
                \pm i \sin{\left( \frac{\alpha_k}{2} \right)}
              \end{array} \right) \: e^{ \pm i k j } 
              \:\:\:\: . 
              \label{exp.10}
              \eneq
\noindent
At variance, to construct solutions with imaginary momentum, we define $\beta_q$ so that

 \beq
\cosh{\beta_q} = \frac{ ( 1 + \gamma ) \left| \sinh   \left( \frac{q}{2} 
\right) \right| }{\sqrt{( 1 + \gamma)^2 \sinh^2 \left( \frac{q}{2} \right) - 
( 1 - \gamma)^2 \cosh^2 \left( \frac{q}{2} \right)}} \;\;\; , \;\; 
\sinh ( \beta_q ) =  \frac{ ( 1 - \gamma)  \cosh \left( \frac{q}{2} \right)
}{\sqrt{( 1 + \gamma)^2 \sinh^2 \left( \frac{q}{2} \right) - 
( 1 -\gamma)^2 \cosh^2 \left( \frac{q}{2} \right)}}
\:\:\:\: . 
\label{exp.11}
\eneq
\noindent
The corresponding solutions are therefore 
given by  

 \beq
\left( \begin{array}{c}
        u_j^E \\ v^E_j 
       \end{array} \right)_{ \pm i  q } = 
 \left( \begin{array}{c}
   \sinh \left( \frac{\beta_q}{2} \right) \\
   \pm \cosh \left( \frac{\beta_q}{2} \right) \ 
        \end{array} \right)
        e^{ \mp q j } 
        \:\:\:\: . 
        \label{exp.12}
        \eneq
        \noindent
In the large-$\ell$ limit, 
the general form of a normalizable, positive-energy solutions will therefore
be given by  
 
 \beq
\left( \begin{array}{c}
        u_j^\epsilon \\ v^\epsilon_j 
       \end{array} \right) = \beta
\sqrt{\frac{2}{\ell}} \Biggl\{ 
\frac{\sinh 
 \left( \frac{\beta_q}{2} \right)}{ \cos \left( \frac{\alpha_k}{2} \right) } \: 
 \left( \begin{array}{c}
                                 \cos \left( \frac{\alpha_k}{2} \right) 
\cos ( k j ) \\ -   \sin \left( \frac{\alpha_k}{2} \right) 
\sin ( k j )                                 
                                \end{array} \right) + 
\frac{\cosh 
 \left( \frac{\beta_q}{2} \right)}{ \sin \left( \frac{\alpha_k}{2} \right) }   \: 
 \left(  \begin{array}{c}
 \cos \left( \frac{\alpha_k}{2} \right) 
\sin ( k j ) \\     \sin \left( \frac{\alpha_k}{2} \right) 
\cos ( k j ) \end{array} \right) - 
 \left( \begin{array}{c}
   \sinh \left( \frac{\beta_q}{2} \right) \\
     \cosh \left( \frac{\beta_q}{2} \right) \ 
        \end{array} \right) 
        e^{ - q j } \Biggr\}  
\:\:\:\: , 
\label{exp.13}
\eneq
\noindent
with $k$ and $q$ related to $\epsilon$ as  from respectively 
Eq.(\ref{exp.6_bis}) and Eq.(\ref{exp.8_cosh})
and  
 
 \beq
\beta = \frac{\sin{\alpha_k}}{ \sqrt{2 \left( \cosh{\beta_q} + 
\cos{\alpha_k} \right) }} 
\:\:\:\: . 
\label{exp.14}
\eneq
\noindent
Finally, from Eqs.(\ref{exp.5}), we see that negative-energy solutions with energy $- \epsilon$ can be recovered 
from the positive-energy ones with energy $\epsilon$ by simply swapping $u_j^E$ and $v_j^E$ with each 
other. (Incidentally, we note that, without resorting to the large-$\ell$ limit, one 
may derive a secular equations for the allowed values of the momentum/energy by 
imposing OBC at both boundary on a general linear combination of all 
four the solutions in Eqs.(\ref{exp.10},\ref{exp.12}). Doing so, one obtains 
that nontrivial solutions are found, provided the 
momenta satisfy the secular equation

\begin{eqnarray}      
\sin^2 \left[ \left( \frac{k - i q  }{ 2 } \right) ( \ell + 1 ) \right] 
[ - 1 + \cos ( \alpha_k + i \beta_q ) ]
+ \sin^2 \left[ \left( \frac{k +  i q  }{ 2 } \right) ( \ell + 1 ) \right] 
[ 1 - \cos ( \alpha_k - i \beta_q ) ] = 0 
 \:\:\:\: . 
 \label{e.6}
\end{eqnarray}
\noindent
Though we have made the exact comparison to the results of \cite{maurizio} only
for specific values of the system parameters, we believe it is safe 
to assume that 
Eq.(\ref{e.6}) is equivalent to the equations of Appendix B of \cite{maurizio}
taken in the specific case $H = J ( 1 + \gamma )$. 

The  formulas we  derived above for the wavefunctions allow us to obtain the explicit 
expressions for  the imaginary time JW-fermion operators at $j=1$, in terms
of which we eventually derive the real fermion operators $\mu_{1 , \lambda} ( \tau ) , 
\mu_{2 ,\lambda } ( \tau)$, entering the derivation of the RG scaling equations for 
the running strengths we performed in section \ref{cross.1}. The relevant 
JW-fermion operators are    given by 

 \begin{eqnarray}
c_1 ( \tau ) &=& \sum_{ \epsilon > 0 } \left[ 
u_1^\epsilon \left( \Gamma_\epsilon + \Gamma_{ - \epsilon }^\dagger \right) 
e^{ -  \epsilon \tau } + 
v_1^\epsilon \left( \Gamma_{ - \epsilon } + \Gamma_\epsilon^\dagger \right)
e^{ \epsilon \tau } \right] 
\nonumber \\
c_1^\dagger  ( \tau ) &=& \sum_{ \epsilon > 0 } \left[ 
u_1^\epsilon \left( \Gamma_{ - \epsilon }  + \Gamma_{  \epsilon }^\dagger \right) 
e^{   \epsilon \tau } + 
v_1^\epsilon \left( \Gamma_{  \epsilon } + \Gamma_{ - \epsilon}^\dagger \right) 
e^{ -  \epsilon \tau } \right] 
\:\:\:\: . 
\label{exp.16}
\end{eqnarray}
\noindent
As for what concerns the corresponding wavefunctions, from the 
explicit calculations one sees that,  for $H =  J ( 1 + \gamma)$
and $\epsilon > 0$, one obtains  the following identities

 \begin{eqnarray}
{\cal A} ( \epsilon ) &=&  u_1^\epsilon + v_1^\epsilon 
= 2 \beta \sqrt{\frac{2}{\ell}} \frac{  \sqrt{\delta^4
 + ( 1 - \delta^2 ) \left( \frac{\epsilon}{J ( 1 + \gamma) } \right)^2}}{ \delta ( 1 + \delta ) }
 e^\frac{\beta_q}{2}   \nonumber \\
 {\cal B} ( \epsilon ) &=&  u_1^\epsilon - v_1^\epsilon 
= 2   \beta \sqrt{\frac{2}{\ell}} \frac{  \sqrt{\delta^4
 + ( 1 - \delta^2 ) \left( \frac{\epsilon}{ J ( 1 + \gamma) } \right)^2}}{ \delta ( 1 - \delta ) }
 e^{-\frac{\beta_q}{2}} 
 \:\:\:\: , 
 \label{exp.15}
\end{eqnarray}
\noindent
with $\delta = ( 1 - \gamma ) / ( 1 + \gamma)$. The solutions with   negative energy 
$- \epsilon$ are simply obtained from 
the ones in Eqs.(\ref{exp.15}) by swapping $u_1^\epsilon$ and $v_1^\epsilon$ with each 
other.  The key quantities we used in section \ref{cross.1} to derive the RG equations 
for the running coupling strengths are the   imaginary-time-ordered Green's functions 
$G_{j , j'} ( \tau )$ for  the real fermions $\mu_{ 1 } ( \tau ) 
= c_1 ( \tau ) + c_1^\dagger ( \tau ) $ and 
$\mu_2 ( \tau ) = -i ( c_1 ( \tau ) - c_1^\dagger ( \tau ) ) $. In terms of 
the functions ${\cal A} ( \epsilon ) , {\cal B} ( \epsilon )$ in Eqs.(\ref{exp.15}), 
one obtains 
 
 \begin{eqnarray}
 G_{1,1} ( \tau ) &=& 2 \sum_{ \epsilon > 0 } {\cal A}^2 ( \epsilon ) 
 \{ [ 1 - f ( \epsilon ) ] {\rm sgn} ( \tau ) e^{ - \epsilon | \tau | } 
 + f ( \epsilon ) {\rm sgn} ( \tau ) e^{   \epsilon | \tau | }  \} \nonumber \\
  G_{2,2} ( \tau ) &=& 2 \sum_{ \epsilon > 0 } {\cal B}^2 ( \epsilon ) 
 \{ [ 1 - f ( \epsilon ) ] {\rm sgn} ( \tau ) e^{ - \epsilon | \tau | } 
 + f ( \epsilon ) {\rm sgn} ( \tau ) e^{   \epsilon | \tau | }  \} \nonumber \\
   G_{1,2} ( \tau ) &=&  - G_{2,1} ( \tau )  = 
  - 2 i  \sum_{ \epsilon > 0 } {\cal A}  ( \epsilon )  {\cal B}  ( \epsilon ) 
 \{ [ 1 - f ( \epsilon ) ] {\rm sgn} ( \tau ) e^{ - \epsilon | \tau | } 
- f ( \epsilon ) {\rm sgn} ( \tau ) e^{   \epsilon | \tau | }  \}
 \:\:\:\: , 
 \label{exp.18}
\end{eqnarray}
\noindent
with $f ( \epsilon )$ being the Fermi distribution function. Incidentally, 
we note  that Eqs.(\ref{exp.18}) describe the real-fermion Green's functions
along all the gapless lines, including the $XX$-lines, provided one uses 
the appropriate expression for ${\cal A} ( \epsilon )$ and ${\cal B} ( \epsilon)$ which,
along this line, is given by

\beq
{\cal A} ( \epsilon ) = {\cal B} ( \epsilon ) = \sqrt{\frac{2}{\ell}}
\: \sqrt{1 - \left[ \frac{\epsilon + H }{2
J} \right]^2}
\:\:\:\:, 
\label{eee.1}
\eneq
\noindent
and sums over energies $\epsilon$ such that $- 2 J - H \leq \epsilon \leq 2 J - H$.
Resorting to  Fourier space to compute $G_{ j , j' } ( i \omega_m ) = \int_0^\beta \:
d \tau \: e^{ i \omega_m \tau } \: G_{ j , j'} ( \tau )$, with $
\omega_m = \frac{2 \pi}{\beta} \left( m + \frac{1}{2} \right)$ being the $m$-th 
fermionic Matsubara frequency, one gets  

 \begin{eqnarray}
 G_{1,1} ( i \omega_m ) &=& -2  \sum_{ \epsilon > 0 } {\cal A}^2 ( \epsilon ) 
 \left( \frac{1}{i \omega_m - \epsilon} + \frac{1}{i \omega_m + \epsilon} \right)
 \nonumber \\
 G_{2,2} ( i \omega_m ) &=& - 2 \sum_{ \epsilon > 0 } {\cal B}^2 ( \epsilon ) 
 \left( \frac{1}{i \omega_m - \epsilon} + \frac{1}{i \omega_m + \epsilon} \right)
 \nonumber \\
 G_{1,2} ( i \omega_m ) &=&  - G_{2,1} ( i \omega_m )= 
- 2 i  \sum_{ \epsilon > 0 }  {\cal A} ( \epsilon ) {\cal B} ( \epsilon ) 
 \left( \frac{1}{i \omega_m - \epsilon} - \frac{1}{i \omega_m + \epsilon} \right)
 \:\:\:\: . 
 \label{exp.20}
\end{eqnarray}
\noindent
In the large-$\ell$ limit, the sum over the momenta in the definition of 
${\cal A} ( \epsilon ) , {\cal B} ( \epsilon )$ can be traded for an 
integral, which, along the CKL1, eventually yields  

 \begin{eqnarray}
  G_{1,1 } ( i \omega_m ) & = & \frac{8}{\pi  J ( 1 + \gamma ) } 
  \: \left( \frac{1-\delta}{1 + \delta}
  \right) \: \int_0^{ 2 J ( 1 + \gamma )  } \: d \epsilon \: \frac{i \omega_m 
  e^{ \beta_q}  \Sigma \left( \frac{\epsilon}{J ( 1 + \gamma) } \right)  }{ \omega_m^2 + \epsilon^2 }  
  \nonumber \\
    G_{2,2 } ( i \omega_m ) & = & \frac{8}{\pi  J ( 1 + \gamma ) } 
    \: \left( \frac{1+ \delta}{1 - \delta}
  \right) \: \int_0^{2 J ( 1 + \gamma )  } \: d \epsilon \: 
  \frac{i \omega_m e^{- \beta_q}  \Sigma \left( \frac{\epsilon}{J ( 1 + \gamma) } \right)  }{ \omega_m^2 + \epsilon^2 }  
  \nonumber \\
   G_{1,2 } ( i \omega_m ) & = &  -  G_{2,1 } ( i \omega_m )
   =   \frac{8}{\pi  J ( 1 + \gamma ) }   \: \int_0^{ 2 J ( 1 + \gamma ) } \: d \epsilon \: 
  \frac{i \epsilon   \Sigma \left( \frac{\epsilon}{J ( 1 + \gamma) } \right)   }{ \omega_m^2 + \epsilon^2 }  
  \:\:\:\: , 
  \label{pp.0}
\end{eqnarray}
\noindent
with the function $\Sigma ( x )$ defined in Eq.(\ref{ccro.7}).
Similarly, along the $XX$-line one obtains 

 \begin{eqnarray}
 G_{1,1} ( i \omega_m ) &=&  G_{2,2} ( i \omega_m ) = 
   \frac{4}{\pi J} \: \int_{-2J - H}^{2J - H} 
 \:  d \epsilon \; \sqrt{1 - \left( \frac{\epsilon + H}{2 J} \right)^2} \: \left[
 \frac{i \omega_m}{   \epsilon^2 + \omega_m^2 } \right] 
 \nonumber \\
  G_{1,2} ( i \omega_m ) &=&  - G_{2,1} ( i \omega_m ) = 
  \frac{4}{\pi J} \: \int_{-2J - H}^{2J - H} 
 \:  d \epsilon \; \sqrt{1 - \left( \frac{\epsilon + H}{2 J} \right)^2} \: \left[
 \frac{i   \epsilon   }{  \epsilon_m^2 + \omega^2 } \right] 
 \:\:\:\: . 
 \label{pertu.9}
\end{eqnarray}
\noindent
 Both Eqs.(\ref{pp.0}) and Eqs.(\ref{pertu.9}) have been used in 
the main text to implement the perturbative expansion in the Kondo-like
Hamiltonian representing the junction. 

To conclude this Appendix, we report the simplified solution available in 
the Ising limit $\gamma = 0$. In this case, from Eqs.(\ref{exp.3}), one
may readily show that the boundary condition at the left-hand endpoint of 
the chain implies $u^E_0 - v^E_0 = 0$, which can be readily satisfied by 
setting 

 \beq
\left( \begin{array}{c}
        u_j^E \\ v_j^E
       \end{array} \right) = \sqrt{\frac{2}{\ell}} 
       \: \left( \begin{array}{c}
                  \cos \left( \frac{\alpha_k}{2} \right) \sin \left( k j - \frac{\alpha_k}{2}
                  \right) \\  - \sin \left( \frac{\alpha_k}{2} \right) \cos 
                  \left( k j - \frac{\alpha_k}{2}
                  \right)  
                 \end{array} \right)
\:\:\:\: . 
\label{exp.21}
\eneq
\noindent
Eq.(\ref{exp.21}), together with the observation that now one has  
$\alpha_k = \frac{\pi}{2} - \frac{k}{2}$ and $\epsilon = 2 J  \sin \left( \frac{k}{2} 
\right)$, implies

 \begin{eqnarray}
 {\cal A} ( \epsilon )&=&  \sqrt{\frac{2}{\ell}} \sqrt{1 - 
 \left( \frac{\epsilon}{2J } \right)^2} \left[ 4 \left( \frac{\epsilon}{2 J} \right)^2 - 1 \right] \nonumber \\
 {\cal B} ( \epsilon ) &=&  2 \sqrt{\frac{2}{\ell}}   \left( \frac{\epsilon}{2J } \right)  \sqrt{1 - 
 \left( \frac{\epsilon}{2J } \right)^2}
 \:\:\:\: . 
 \label{exp.22}
\end{eqnarray}
\noindent

\biboptions{sort}

\end{document}